%% file: main.tex
\documentclass[twocolumn,tighten,resetfootnote]{aastex701} 
\usepackage{enumitem}
\usepackage{amsmath}
\usepackage[dvipsnames]{xcolor}

\usepackage[dvipsnames]{xcolor}
\newcommand\refedit[1]{#1}
\newcommand\newrefedit[1]{#1}

\graphicspath{{./}{figures/}}
\usepackage{amsmath}
\usepackage{graphicx}
\usepackage{float}
\usepackage{caption}

\begin{document}

\title{Variations in the 3.3~$\micron$ Polycyclic Aromatic Hydrocarbon Feature Across Nearby Galaxies Driven by Metallicity and Radiation Field Spectrum}
\shorttitle{3.3 $\micron$ PAH Feature in Nearby Galaxies}
\shortauthors{Koziol et al.}
\correspondingauthor{Hannah Koziol}
\email{hkoziol@ucsd.edu}
\input{affiliations}
\input{authors}

\begin{abstract}
We use JWST NIRCam imaging to investigate the 3.3~$\micron$ polycyclic aromatic hydrocarbon (PAH) feature in nearby galaxies. NIRCam observations of the 3.3~$\micron$ feature are emerging as a powerful tool for studying the structure of the interstellar medium (ISM) and the conditions of the dust at $\sim$0\farcs1 resolution. These maps require accurate subtraction of the underlying continuum emission. We present an empirical method to isolate the PAH-correlated emission in the F335M filter using the F300M and F360M filters for continuum subtraction. We find that the slope of the F335M/F300M versus F360M/F300M colors for PAH-correlated emission shows a dependence on local ISM properties, with the strongest dependence on specific star formation rate. Weaker emission features captured by these bands appear suppressed relative to the main 3.3~$\micron$ feature in regions of active star formation. We find trends in the 3.3/7.7 and 3.3/11.3~$\micron$ ratios that suggest changes in PAH size, charge, and heating by a varying radiation field spectrum. We find decreases in both band ratios with increasing metallicity, which we attribute to a shift to smaller PAH populations at low metallicity. Comparison to optical ionized gas line ratios and dust models show that variations in the interstellar radiation field spectrum influence the PAH feature ratios. This analysis supports inhibited growth formation scenarios for the observed PAH band ratio trends with metallicity and emphasizes the importance of considering the local radiation field characteristics and gas-phase metallicity when using these band ratios as PAH property diagnostics. 

\end{abstract}

\keywords{Polycyclic Aromatic Hydrocarbons (1280) --- Interstellar Medium (847) --- \newline Extragalactic Astronomy (506) --- Interstellar Dust (836)}

\section{Introduction} \label{sec:intro}
Polycyclic aromatic hydrocarbons (PAHs) are small dust grains composed of aromatic carbon rings with hydrogen atoms attached that emit in specific vibrational bands in the infrared, the strongest of these being centered at 3.3, 6.2, 7.7, 8.6, 11.3, 12.7, and 17~$\micron$ \citep{Allamandola1989,Smith2007}. PAHs can be heated sufficiently by the absorption of a single UV photon to emit in the mid-IR \citep{Sellgren1984}. They are ubiquitous in the ISM of the Milky Way, making up $\sim$5\% of its total dust mass, and can contribute up to 20\% of the total IR emission observed from galaxies \citep{Draine2007, Smith2007}. Photoelectric heating from PAHs influences the phase structure of the ISM \citep{Hollenbach1999, Wolfire2003}, and models and observations show that the smallest, neutral PAHs contribute the most to heating \citep{Bakes1994, Tielens2008}. For this reason, measuring the PAH size and charge distribution is crucial for understanding the ISM and energy budget of galaxies. 

In addition, recent JWST studies have shown a strong correlation between PAH emission and molecular gas surface density \citep{Leroy2023}. The 3.3~$\micron$ feature has been shown to have a tight correlation with molecular gas, covering more than two orders of magnitude in CO and PAH intensities \citep{Chown2025a}. The high angular resolution of NIRCam imaging at 3.3~$\micron$\ ($\sim0\farcs1$) makes this PAH feature one of the highest angular resolution tracers of the cold ISM. \refedit{It has also been used as a tracer of dust-embedded star clusters in nearby galaxies \citep{Rodriguez2023, Graham2025}.}
Observations of NGC\,628 from the Feedback in Emerging Extragalactic Star Clusters survey revealed a tight sublinear correlation between 3.3~$\micron$ PAH emission and SFR traced by Br$\alpha$ around emerging young star clusters \citep{Gregg2024}.

The average characteristics of PAH populations, including size, charge, and structure, influence which features they emit \citep{Draine2007, Tielens2008}. The vast majority of 3.3~$\micron$ emission is from small PAHs ($\lessapprox$ 100 C atoms), while larger PAHs ($\sim 100 - 1,000$ C atoms) contribute mostly to longer wavelength PAH bands such as the 11.3~$\micron$ feature \citep{Desert1990, Schutte1993, Draine2021}. Single photon absorption allows small PAHs to reach a temperature high enough to radiate in the short wavelength features, including the 3.3~$\micron$ feature, while larger PAHs reach lower temperatures \citep{Schutte1993}. The 3.3 and 11.3~$\micron$ features are dominated by C-H bonds in neutral grains, and the 6.2 and 7.7~$\micron$ C-C features largely come from ionized PAHs \citep{vanDiedenhoven2004, Draine2007, Boersma2016}. For this reason, the ratio of the 3.3/11.3~$\micron$ PAH features is used as a tracer of PAH size, and the ratios of the 3.3/7.7 and 7.7/11.3~$\micron$ PAH features trace a combination of size and charge \citep[e.g.,][]{Allamandola1999, Draine2001, Maragkoudakis2020, Rigopoulou2021}. 

\refedit{In the following, we use PAH band ratios to trace the evolution of grain size and charge in the ISM. However, r}ecent work has shown that these band ratios can also be affected by the spectrum of the radiation field heating the PAHs \refedit{\citep{Maragkoudakis2020, Draine2021, dale2023, Baron2024, dale2025}}. Harder radiation fields with higher average photon energies result in hotter PAHs, which lead to increased emission at shorter wavelengths relative to longer wavelengths. Changes in PAH sizes and changes in radiation field hardness behave similarly in 3.3/11.3 and 7.7/11.3~$\micron$ band ratios because both of these effects change the temperature of the PAHs. \citet{Chastenet2023} found trends in the 3.3/11.3~$\micron$ ratio that could be explained by a combination of changes in the PAH size distribution and the spectrum of the radiation field heating the PAHs. \citet{Baron2024} identified strong correlations between PAH band ratios and optical ionized gas line ratios in three nearby galaxies, \refedit{which they interpret as being due to radiation field spectrum effects}. \citet{Baron2025} then confirmed that the relation between the 7.7/11.3~$\micron$ PAH band ratio and optical line ratios (e.g., [SII]/H$\alpha$) is a universal feature across nearby star-forming massive galaxies on 40--150 pc scales. They interpret these trends as the ionized gas and PAHs being exposed to two different parts of the same \refedit{spectrally varying} radiation field. The ionization of the warm ionized gas is set by the spectrum $\leq$ 912\AA, while the PAHs are heated by non-ionizing photons.

In addition to radiation field hardness effects, there are trends in the observed band ratios that suggest evolution of PAH size and charge with environment. One dramatic example is the behavior of PAH emission in the vicinity of active galactic nuclei (AGN).  \citet{Smith2007} demonstrated that the 7.7~\micron\ feature is often weak in the vicinity of AGN relative to the 11.3~\micron\ feature. This effect has also been observed in several Seyfert galaxies \citep{DiamondStanic2010, Garcia-Bernete2022, Zhang2024} and other AGN \citep[e.g., in NGC\,4138;][]{Donnelly2024}. The Galactic Activity, Torus, and Outflow Survey (GATOS) survey observed that in Seyfert-like AGNs, the ionization state of the PAHs can be affected out to kpc scales \citep{GarciaBernete2024}. Using JWST observations including the 3.3~\micron\ feature, \citet{Lai2023} also found distinct differences in the PAH band ratios in star-forming regions and in the vicinity of the AGN in NGC\,7469. \refedit{Increased relative 11.3~$\micron$ feature strength compared to 3.3 and 7.7~$\micron$ has been interpreted as the destruction of small and ionized PAHs in the hard radiation field produced by the AGN} \newrefedit{\citep{Garcia-Bernete2022, GarciaBernete2024}.} \refedit{Alternatively, \citet{Donnelly2024} interpreted the relative increase being driven by a soft FUV field produced by an older stellar population surrounding the AGN and not a shift to a more neutral PAH population.}

\refedit{Another example of PAH band ratio trends that appear to be driven by charge and size is the behavior of PAHs as a function of metallicity.} Shorter wavelength PAH features are observed to be more prominent relative to longer wavelength PAH features at low metallicity \citep{Smith2007, Sandstrom2012, Whitcomb2024}. This difference was first observed with {\em Spitzer} as a trend of weaker 17~$\micron$ feature strength compared to shorter wavelength PAH features and higher 6.2/7.7~$\micron$ ratio at low metallicity. With JWST, this trend has also been observed as an increase in the relative strength of the 3.3~$\micron$ feature (\citealp{Lai2020, Lai2025, Whitcomb2025}). \citet{Whitcomb2025} showed that the observed shift to higher 3.3/11.3~$\micron$ ratio at lower metallicities in M101 could not be explained by changes in the radiation field spectrum alone, and reflects a shift towards smaller PAH sizes. A potential explanation for this shift is the low abundance of carbon in low-metallicity environments, impeding the creation of larger PAHs, causing the ratio of small to large PAHs to increase. \refedit{Using their ``inhibited growth'' model, \citet{Whitcomb2024} replicated observed band ratios in low metallicity environments by decreasing the average PAH size and PAH fraction below a certain threshold metallicity value.} 

While the 3.3~$\micron$ feature is typically the most prominent near-infrared (NIR) PAH feature, the 3--3.5~$\micron$ wavelength range includes several other emission features from small carbonaceous dust grains, including the 3.4~$\micron$ aliphatic feature, 3.47~$\micron$ plateau feature, and PAH ``continuum''. The 3.4 $\micron$ feature is attributed to a C-H stretching mode of aliphatic bonds \citep{Pendleton2002} with a typical ratio of aliphatic-to-aromatic feature strengths being $\sim$10\% \citep{Lai2020} when measured using spectral decomposition tools such as PAHFIT \citep{Smith2007}. Aliphatic bonds are more fragile, meaning a higher aliphatic content may indicate more recent PAH formation, providing insight into the beginning of the life cycle of PAHs \citep{Yang2017, Hammonds2015}. The 3.47~$\micron$ plateau is thought to be caused by super-hydrogenated PAHs \citep{Hammonds2015, Lai2020}. Recent JWST observations have also revealed a continuum, attributed to PAHs, beginning at $\sim$1~$\micron$, sharply increasing at about 3.2~$\micron$, and continuing to rise until $\sim$ 5~$\micron$ that has been characterized in NIRSpec observations of seven nearby objects along the low mass stellar life cycle \citep{Boersma2023}. This continuum is correlated with PAH emission, but it is not currently known what specific carriers contribute to it. It is possible that overtone and combination bands from PAHs are part of it \citep{Boersma2023, Peeters2024}.

{\em Spitzer} Space Telescope observations of PAHs within the 5--20~$\micron$ range contributed greatly to advancement in our understanding of PAHs \citep{Li2020}, but its spectroscopic instrument did not cover the 3.3~$\micron$ regime and its wide 3.6~$\micron$ photometric filter is highly dominated by starlight. 
Because this feature is difficult to observe from the ground, space-based missions including ISO and {\em AKARI} provided key measurements, but it could only be observed in the brightest regions in the Milky Way or starburst galaxies \citep{Sloan1997, Verstraete2001, Lai2020}. The 3.4~$\micron$ aliphatic feature and the faint, broad 3.47~$\micron$ plateau were observed with ISO and {\em AKARI} in integrated spectra of bright galaxies \citep{Lai2020}. With JWST, we now have the opportunity to study the 3.3~$\micron$ PAH and neighboring features throughout the ISM of normal, nearby galaxies at much higher angular resolution with NIRCam ($\sim$0\farcs10-0\farcs12 at F300M, F335M, and F360M) and over large fields of view (9.7 square arcminutes).

In order to isolate the 3.3~$\micron$ PAH emission, we must first remove the stellar continuum and other sources of emission from the F335M \citep{Lai2020}. Without the ability to spectrally decompose, the surrounding bands (F300M and F360M) become crucial to extracting the PAH emission in F335M. In the following, we use a method initially presented by \citet{Sandstrom2023}, building on \citet{Lai2020}, which uses empirically determined F300M/F335M/F360M colors to isolate all PAH-related emission in F335M. Specifically, in this work, we optimize the color-color-based subtraction presented by \citet{Sandstrom2023} using the F300M, F335M, and F360M NIRCam filters on 19 galaxies in the Physics at High Angular resolution in Nearby Galaxies (PHANGS) survey. 

The remainder of this paper is organized as follows: Section \ref{sec:data} describes the observations and data products used in this work. Section \ref{sec:method} describes the new prescription developed for mapping the PAH emission in the F335M filter and what drives the variation in the features in the 3--3.5~$\micron$ region. In Section \ref{sec:results} we investigate the variation of the PAH properties with environment using the resulting continuum-subtracted F335M maps relative to other PAH features, including the 7.7 and 11.3~$\micron$ features traced with MIRI photometry. Section \ref{sec:discussion} discusses the implications of these results for the properties of PAHs in the context of ISM conditions and galaxy properties. Finally, Section \ref{sec:conclusion} summarizes and concludes the key results of this work. 

\section{Data} \label{sec:data}
\input{table_phangs_sample}
We use observations obtained as part of the PHANGS survey \citep{Leroy2021}. We use JWST observations of 19 galaxies from the PHANGS-JWST Cycle 1 Treasury (GO 2107; PI Lee) presented by \citet{Lee2023} and \citet{williams2024}. Galaxy properties are listed in Table \ref{tab:table}. The sample includes galaxies from the PHANGS-ALMA \citep{Leroy2021} and PHANGS-MUSE \citep{emsellem2022} surveys with inclinations lower than 60$^{\circ}$ and distances between 5--20~Mpc. Details on the JWST imaging reduction pipeline can be found in \citet{williams2024}.

\subsection{JWST Photometry}
We use an updated reduction of the PHANGS--JWST Cycle 1 Treasury data presented by \citet{Chown2025a}. The PHANGS-JWST Cycle 1 galaxies were observed in eight photometric filters with NIRCam and MIRI covering a wavelength range from 2--21~$\micron$. The primary filters used in this work include the NIRCam filters F300M, F335M, and F360M for mapping the 3.3~$\micron$ feature. The F335M maps for our sample have physical resolutions between 3--12~pc (corresponding to the 0\farcs1 angular resolution at the distance of the targets). 

In our analysis, we also use observations from two MIRI bands dominated by PAH emission: F770W and F1130W. \refedit{In our analysis of the $\sim$3~$\micron$ features, we compare with photometrically traced 7.7 and 11.3~$\micron$ emission observed with the F770W and F1130W filters. These filters include both PAH emission, dust continuum, and starlight. Generally, the starlight is much fainter than the PAH emission and dust continuum emission, but there are places with low ISM content but high stellar mass surface density in our galaxies where starlight subtraction is needed at 7.7 \micron. We subtract stellar continuum from F770W using the method presented by \citet{Sutter2024}. Following a similar analysis performed by \citet{Helou2004} using {\em Spitzer} 3.6~$\micron$ and 8~$\micron$, \citet{Sutter2024} scaled the F300M data to predict the amount of starlight contaminating F770W using Code for Investigating GALaxy Emission \citep[CIGALE;][]{boquien2019} models for a reasonable range of stellar populations. For the remainder of this paper, we use the stellar-subtracted F770W, denoted by F770W$_{\rm ss}$. The F1130W filter is used to measure the 11.3~$\micron$ PAH feature, and the stellar continuum is negligible at this wavelength. We do not remove dust continuum from the filters. Hands et al. (in prep.) and \citet{Whitcomb2023} shows this is a $<$20\% contribution to the filters.}

\subsection{Ancillary PHANGS Data}
To measure local environmental properties within the galaxies, including specific star formation rate (sSFR), we use smaller regions with associated physical properties presented by \citet{sun2022} and \citet{sun2023}. We use properties calculated in 1.5~kpc hexagonal and 500~pc deprojected galactocentric radial annular regions, including surface densities of atomic and molecular gas, stellar mass, and star formation rate. This multi-wavelength collection of data provides a wealth of information we use to calculate different properties on small scales across the galaxies to compare to measurements made in this analysis. 

We use IFU data from the Multi Unit Spectroscopic Explorer (MUSE) on the Very Large Telescope (VLT) in the PHANGS-MUSE survey presented by \citet{emsellem2022} to measure optical line ratios to investigate PAH band ratio dependence on radiation field hardness. This survey covered the 19 galaxies used in this sample at $\sim$1$\farcs$0. For this analysis, we use the maps convolved to an angular resolution of 150 pc, with spaxels that correspond to a size of 150~pc to increase the signal-to-noise ratio for comparison to the PAH maps. In this analysis, following \citet{Baron2024}, we use the H$\alpha$ and H$\beta$ Balmer lines in combination with the [OIII]5007\AA, [NII]6584\AA, and [SII]6717\AA + 6731\AA~lines.

\refedit{We use HII region based gas-phase 12+log(O/H) maps of the 19 JWST Cycle 1 galaxies to measure the correlation between PAH properties and metallicity. \citet{Williams2022} used data from the PHANGS-MUSE survey to measure the [NII], [SII], [OIII], H$\alpha$, and H$\beta$ intensities in HII regions and used Gaussian Process Regression (GPR) to create resolved metallicity maps that smoothly interpolate the measured HII region metallicities across the galaxy disk. They found the difference between GPR interpolated maps and radial gradients from HII regions to be small. The 12+log(O/H) measurements are based on the Scal prescription presented by \citet{Pilyugin2016}.}

\subsection{Data Processing}\label{sec:processing}
\refedit{To ensure matched angular resolution for the resolved analysis}, the F300M, F335M, and F360M filters are first convolved with the PSF of the other two filters to create uniform resolution across the images\refedit{, at the expense of a small amount of spatial resolution for smooth subtraction}. \refedit{We use PSFs of the three medium bands generated from \texttt{STPSF} \citep{Perrin2014} and convolve each filter with the PSF of the other two bands.} We call the final PSF \refedit{that results from convolving} the three NIRCam filters the `triple cross-convolved PSF'. We do not create PSF-matching kernels between the \refedit{$\sim$3~\micron} filters, as kernel creation becomes unstable when the input and target PSFs are too similar. \refedit{In addition}, the PSFs are not circularized. 

For comparison with longer wavelength PAH features, we generate kernels to convolve from the triple cross-convolved PSF of the continuum subtracted F335M maps to the F1130W resolution following the method presented by \citet{Aniano2011}, where these PSFs are circularized in the process of kernel creation. We convolve and regrid the continuum-subtracted F335M and F770W maps to the resolution and native F1130W pixel scale using the same convolution method and the \texttt{Astropy} \texttt{reproject} function. For comparison to the MUSE data, we \refedit{convolve} the PAH maps \refedit{to} the spatial scale of 150~pc using kernels individually generated to go from their native resolution to a Gaussian with the corresponding angular resolution, depending on the galaxy's distance. 

Our NIRCam data have visible striping patterns due to ``1/$f$ noise''. This noise pattern arises from fluctuations in the reference voltage of the readout amplifiers of NIRCam. This results in stripes of additive offsets along the readout direction that are correlated with time and well described by a 1/$f$ power spectrum, with typical values around $\pm$ 0.05~MJy~sr$^{-1}$ \citep{Schlawin2020}. We mask data below the typical stripe value in each galaxy, visually inspected to be $\pm$0.04~MJy~sr$^{-1}$, in the continuum-subtracted F335M maps. To select parts of the map with well-detected PAH emission, we also use a mask based on the F1130W maps, because the signal-to-noise in those maps is usually high relative to any background issues \refedit{in the F335M$_{\rm{PAH}}$ maps}. We make a 3$\sigma$ noise cut in the F1130W filter, then make a mask to include all pixels above the 50th percentile intensity in the remaining S/N $>$ 3 F1130W data. The 3$\sigma$ value was determined by \citet{Sutter2024} by using regions of empty sky in the F1130W maps that included areas off the galaxy. In each galaxy, an average value of 16\%, with a range between 7-32\%, of the \refedit{F335M$_{\rm{PAH}}$} flux remains after these cuts. We apply this cut to the continuum-subtracted F335M and star-subtracted F770W images to isolate pixels with PAH emission. More details on the image processing can be found in Appendix \ref{sec:more_processing}.
  
\section{F335M Continuum Removal} \label{sec:method}
\subsection{Previous Work on F335M Continuum Removal}
Several methods for isolating the 3.3~$\micron$ feature have been published, each making different assumptions about the shape of the underlying continuum and whether the method isolates purely 3.3~\micron\ PAH or all PAH-related emission\refedit{, also including part of the 3.4~$\micron$ feature, 3.47~$\micron$ plateau, and PAH-correlated continuum,} in the F335M filter. \citet{m82} used a combination of the F250M and F360M to scale and remove the continuum in observations of M82, and \citet{Gregg2024} applied a linear interpolation in the surrounding F277W and F444W in observations of star clusters, both including contributions from other PAH-correlated emission in F335M. In Sextans A, \citet{Tarantino2025} used the flanking filters F300M and F360M for F335M continuum subtraction by assuming the F300M has no PAH contamination, the F360M PAH contamination scales linearly with the PAH emission in the F335M filter, and the continuum slope is linear between the F300M and F360M filters. Based on the spectral decomposition of the data presented by \citet{Chown2025}, \citet{Tarantino2025} calculated a proportionality constant to adjust the scale of the PAH contamination in F360M to only include the 3.3~$\micron$ PAH feature. \citet{Whitcomb2025} employ a combination of several methods for removing the continuum, including the synthetic photometry-based prescription from \citet{Lai2020} and the method in \citet{Tarantino2025}, both spectroscopically calibrated to only include the 3.3~$\micron$ feature. 

\subsection{Our Method}
Our approach is based on the method outlined by \citet{Lai2020}, which showed that the slope of the underlying continuum emission between the F335M and F360M filters is not fixed. Using {\em AKARI} and {\em Spitzer} spectra to predict synthetic JWST NIRCam photometry, \citet{Lai2020} isolated the 3.3~$\micron$ PAH feature for extraction. Their continuum color can be described by 
 \begin{equation} \label{lai}
     \frac{\mathrm{F335M}_{\rm cont}}{\mathrm{F300M}} = A_{L20} + B_{L20}\frac{\mathrm{F360M}}{\mathrm{F300M}}
 \end{equation}
with values $A_{L20} = 0.35$ and $B_{L20} = 0.65$ with $F$ representing flux density, $F_\nu$. \citet{Sandstrom2023} found that with high angular resolution data, there is diffuse emission in the F360M filter correlated with PAH emission, which may complicate the use of F360M as a continuum tracer. Their method instead isolates all PAH-correlated emission in F335M, which includes contributions from other features and the PAH  continuum, and corrects for PAH contamination in the F360M filter. To do this, \citet{Sandstrom2023} modified the \citet{Lai2020} procedure to measure the PAH contamination in the F360M filter using the observed F335M/F300M and F360M/F300M colors in PAH-bright regions where F1130W, a PAH-dominated filter with negligible stellar continuum, is greater than 10~MJy sr$^{-1}$, 
using 
 \begin{equation} \label{s23}
     \frac{\mathrm{F335M}_{\rm PAH}}{\mathrm{F300M}} = A_{S23} + B_{S23} \frac{\mathrm{F360M}}{\mathrm{F300M}}
 \end{equation}
 with $A_{S23} = -0.2$ and $B_{S23} = 1.6$. In the case where the slope $B_{23}$ is fixed, the difference in the \citet{Sandstrom2023} and \citet{Lai2020} methods is a constant offset, as described in \citet{Whitcomb2025}. 
 
At each point in the maps, the measured $x_m =$ F360M/F300M and $y_m=$ F335M/F300M values are scaled to intersect with the \citet{Lai2020} continuum equation to obtain corrected F360M/F300M ($x_c$) and F335M/F300M ($y_c$) values. 
 Equation 9 in \citet{Sandstrom2023} gives 
 \begin{equation}
     x_c = \frac{B_{S23}x_m-y_m+A_{L20}}{B_{S23} - B_{L20}}
 \end{equation}
 and $y_c$ is given by 
 \begin{equation}
     y_c = B_{L20} (x_c) + A_{L20}.
 \end{equation}
 The corrected continuum can then be calculated with
 \begin{equation}
 \mathrm{F335M}_{\rm cont} = y_c \cdot \mathrm{F300M}
 \end{equation}
 and the corresponding PAH emission in the F335M filter can be found with 
 \begin{equation}
     \mathrm{F335M}_{\rm PAH} = \mathrm{F335M} - \mathrm{F335M}_{\rm cont}.
 \end{equation}
 
 Our work includes four improvements that build upon prior analyses: 1) the PSF is matched across filters to better account for small resolution changes between the three bands, 2) the color cuts are fine-tuned to more accurately remove stellar continuum in the \refedit{$B_{S23}$} measurements, 3) the slope of the PAH\refedit{-correlated}
 emission \refedit{in F335M and F360M} is calculated individually in 1.5~kpc regions in each galaxy and across each galaxy as a whole and 4) we cover a large sample of 19 galaxies with varying environmental conditions. In the following, we demonstrate that the slope \refedit{B$_{S23}$, which we refer to as B$_{\rm{PAH}}$,} systematically varies with galactic environment.

\begin{figure*}[ht]
      \centering
      \includegraphics[width = \textwidth]{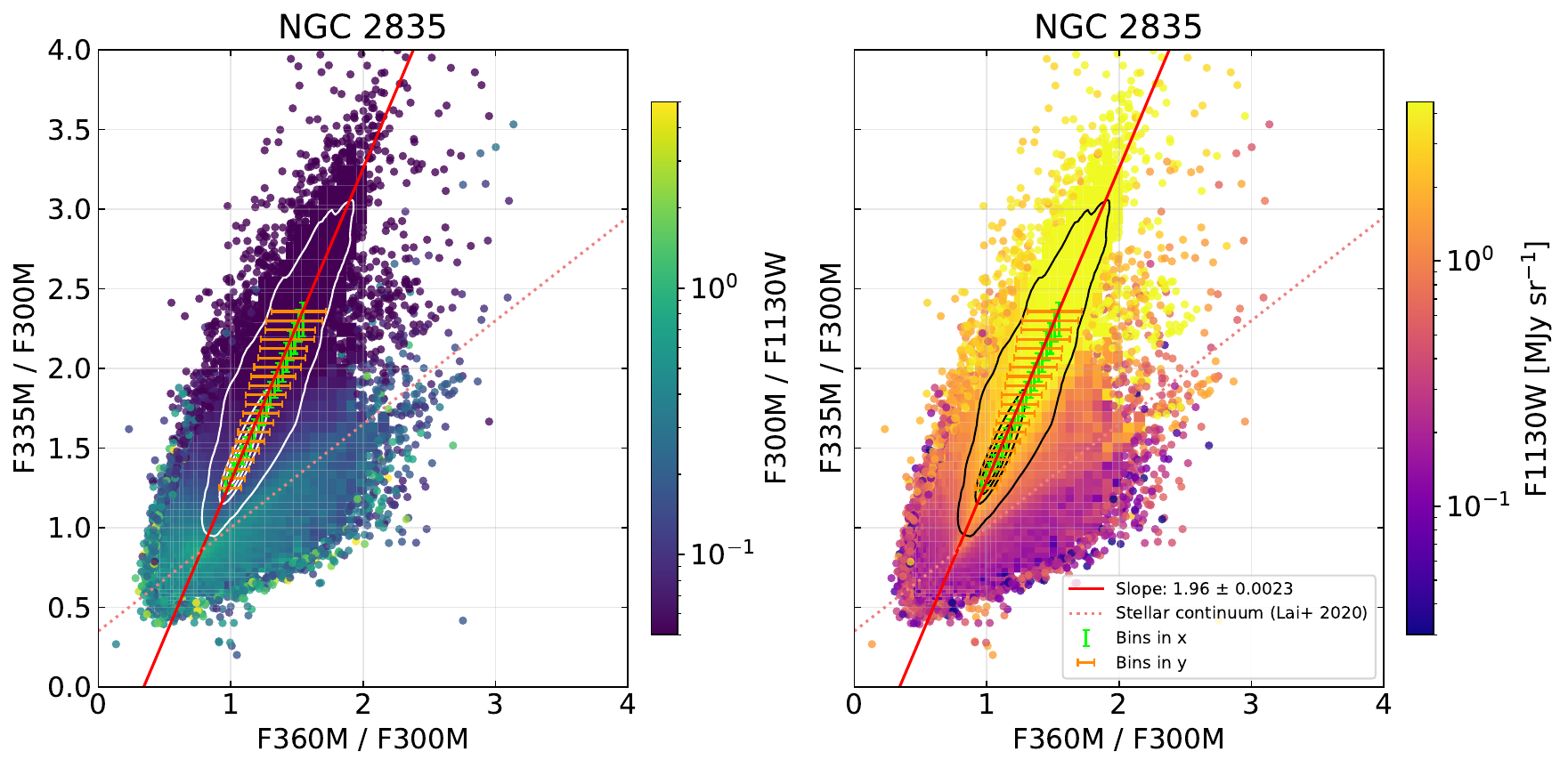}
      \caption{Example slope ($B_{\rm PAH}$) measurement for galaxy NGC\,2835 \refedit{shown in red}. All pixels in the F335M/F300M and F360M/F300M color space are shown in a 2D histogram \refedit{where the bin edges are logarithmically spaced in both dimensions}. Bins with fewer than 10 pixels are shown as individually scattered points. Only the pixels meeting the criteria to be PAH-dominated, F1130W $>$ 3~MJy~sr$^{-1}$ and F300M/F1130W $<$ 0.1, are used for the binned medians described in Section~\ref{sec:B_pah_phangs}. \refedit{The contours enclose where the 90th percentile and the highest density of the data lie.}
      The green bars show the data binned along the $x$-axis, while the orange bars are binned along the $y$-axis with 16th-84th percentile error bars shown. The bisector slope between the binned medians is fit by the red solid line \refedit{with errors on the slope calculated using the results from our forward model} and the dotted line shows the stellar continuum slope from \citet{Lai2020}.}
      \label{fig:color_cuts}
  \end{figure*}

\subsection{Measurements of $B_{\rm PAH}$ in PHANGS Galaxies} \label{sec:B_pah_phangs}
 \begin{figure*}[ht]
      \centering
      \includegraphics[width = \textwidth]{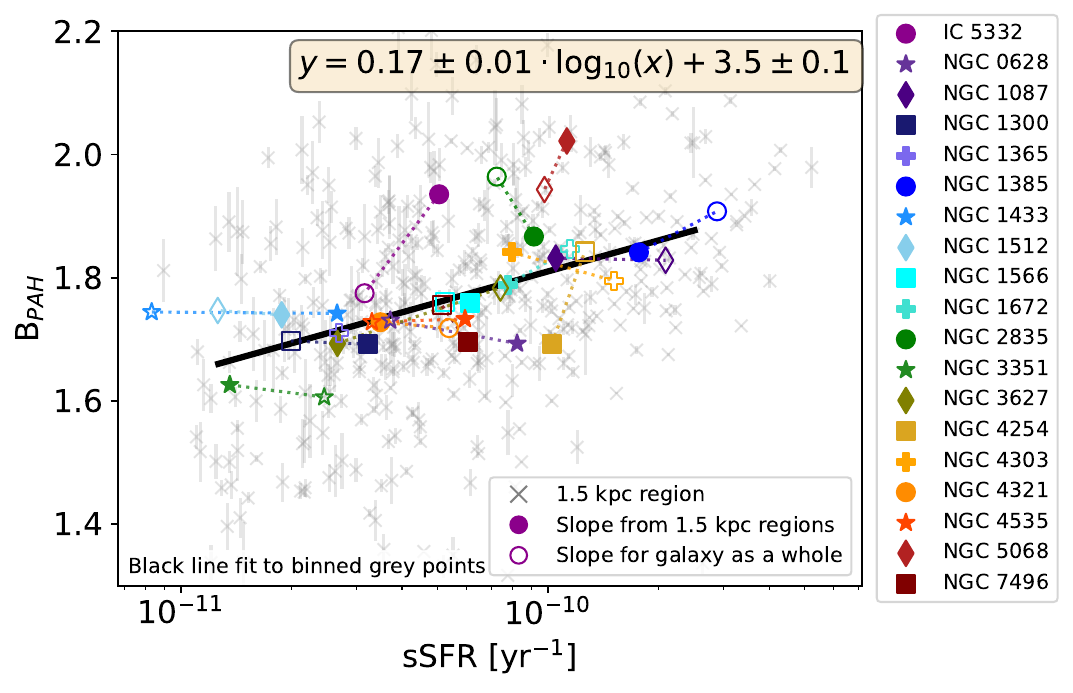}
      \caption{Slope values as a function of sSFR for all 1.5~kpc regions evenly sampled across all galaxies. The slope values were calculated individually in each region, following the same procedure from Section \ref{sec:method}. The best fitting line, $B_{\rm PAH} = 0.17 \pm 0.01 \times \log_{10}({\rm sSFR} / {\rm yr^{-1}})
      +3.5 \pm 0.1$, 
      was calculated on binned medians and overplotted in black. Average slope and sSFR values from regions inside each galaxy are plotted as various shapes and solid colors. Galaxy property values and regions are from \citet{Sun2020}. Slope values measured from treating the entire galaxy as one aperture are shown in the same shapes and colors with open faces. These are plotted with the global average sSFR value in the JWST coverage for each galaxy. Galaxy-wide and region averaged slope measurements are the same in some galaxies, while the effects of weighting each region equally, regardless of number of points, can offset them in other galaxies.}
      \label{fig:megatable_regions}
      
      \end{figure*}

  \begin{figure}[ht]
      \centering
      \includegraphics[width = \columnwidth]{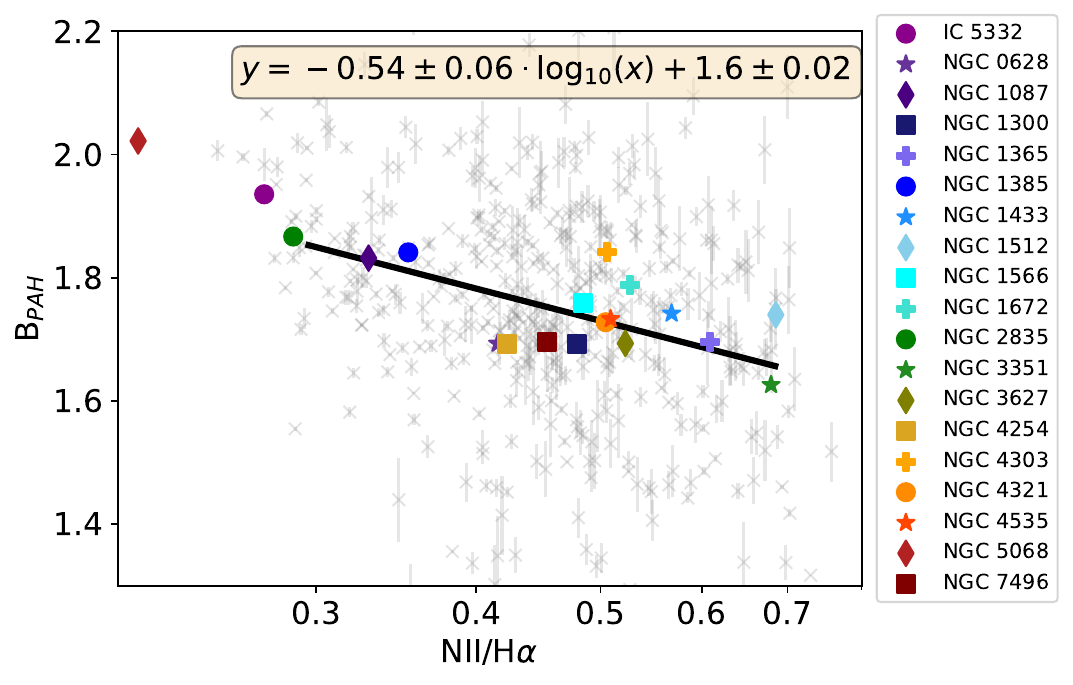}
      \caption{Slope as a function of optical line ratio [NII]/H$\alpha$. The grey and filled points represent the same information as Figure~\ref{fig:megatable_regions}, but with [NII]/H$\alpha$ on the x-axis.}
      \label{fig:megatable_regions_optical}
  \end{figure} 
We fit the slope of the F335M/F300M vs F360M/F300M colors of PAH-dominated emission to get our new values of $B_{S23}$, which we refer to as $B_{\rm PAH}$, both in 1.5~kpc regions and across each PHANGS-JWST Cycle 1 galaxy as a whole, at the resolution of the F1130W data.  \citet{Sandstrom2023} determined the PAH-correlated emission slope using a simple surface brightness cut, including all regions, where F1130W~$>$~10 ~MJy~sr$^{-1}$ in NGC\,628, NGC\,7496, and IC\,5332. With the wider range of targets spanned in this paper, we find that this cut is not sufficient to isolate emission that is highly PAH dominated. In galaxies with bright starlight, mostly at their centers, the PAH-dominated regions are not well isolated by the F1130W-only based cut. To account for this we instead use a new color cut shown in Figure \ref{fig:color_cuts}, F300M/F1130W $<$ 0.1, to isolate pixels where the stellar-dominated F300M is small compared to the PAH-dominated F1130W. With this new color cut, the cut in F1130W did not need to remain as strict---we changed this cut to F1130W $>$ 3~MJy~sr$^{-1}$.

Having selected the subset of emission from highly PAH-dominated regions with the F300M/F1130W cut, we proceed to determine the slope in color-color space, $B_{\rm PAH}$.  We first \refedit{measured} binned medians of the PAH-dominated points into 20 bins evenly spaced between the 5th and 95th percentiles of the data and use these binned points to produce a fit in the F335M/F300M vs F360M/F300M color space. Because there are comparable uncertainties in both the $x$ and $y$ directions, we bin the data in both directions, perform linear regressions on each set of bins, and use the bisector of the two fits as our best-fit relation \citep{Isobe1990}. 

To test our recovery of the $B_{\rm PAH}$ slope and quantify uncertainties in our fitting procedure, we forward model the observations with realistic starlight and PAH contributions. Model F335M data \refedit{are} created \refedit{from the F1130W maps} using the average 3.3/11.3~$\micron$ PAH ratio per galaxy for the PAH contribution and adding 0.88$\times$F300M for the starlight contribution; the coefficient \refedit{for which} is measured in Appendix \ref{sec:without_360}. Modeled F360M data \refedit{are} created assuming a value for $\rm B_{\text{PAH}}$\refedit{, which determines the PAH contribution to F360M relative to F1130W. The stellar continuum components of the modeled F335M and F360M are then chosen to satisfy the Equation \ref{lai} by \citet{Lai2020}.}
\refedit{For each galaxy or region, we generate 100 modeled data sets with random errors using the measured $\rm B_{\text{PAH}}$ as an input slope. The slope is then remeasured following the procedure described above. The \refedit{median} difference between the input and recovered slopes is used as a measure of the bias, which is then used to correct the measured slope. We report the bias-corrected $\rm B_{\text{PAH}}$ and the standard deviation of the resulting slopes from the trials as the error.}
Figure \ref{fig:color_cuts} shows an example \refedit{result of} this procedure for the galaxy NGC\,2835. The stellar continuum slope from \citet{Lai2020} is plotted for comparison to show that the updated selection criteria isolate the PAH-dominated lines of sight.

To investigate the dependence of the slope on various galaxy properties, we divided each galaxy into the 1.5~kpc regions presented by \citet{sun2022}, and recalculated the slopes in each region. We only considered regions that are at least 50\% filled by the JWST coverage; then we applied the same color-cuts \refedit{described in the beginning of this section} and only included regions with a sufficient number of points after these cuts, at least 25 points where F300M/F1130W $<$ 0.1 and F1130W $>$ 3~MJy~sr$^{-1}$, in the analysis. The 1.5~kpc regions are fit in the same way as galaxy-wide measurements. The errors on these measurements are calculated using the same forward modeling technique within the regions.

\subsection{$B_{\rm PAH}$ Variation}\label{sec:bpah_meas}
We observe variations of the slope values between galaxies and a systematic change on 1.5~kpc scales with environment within each galaxy. When treating each galaxy as a single aperture and making one $B_{\rm PAH}$ measurement, we find a median value of 1.76 with a 16th-84th percentile range of 1.72-1.85 among the 19 galaxies in our sample. When measured on 1.5~kpc scales, we measure a median value of 1.76 and an 16th-84th percentile range of 1.57-1.92.

We test the correlation of the measured slope in the 1.5~kpc regions with several different galactic environmental properties, all shown in Table \ref{tab:correlations}, and find that sSFR and the optical line ratio [NII]/H$\alpha$ have the strongest correlations. We calculated sSFR using SFR surface density measured with Balmer-decrement corrected H$\alpha$ measurements \citep{Belfiore2022} divided by the stellar mass surface density calculated with Near-IR {\em Spitzer} IRAC 3.6~$\micron$ and WISE1 data \refedit{\citep{Leroy2021}}\refedit{, where we average the values in each region, then take the ratio}. The optical line ratios were averaged in each 1.5~kpc region. 
Figures \ref{fig:megatable_regions} and \ref{fig:megatable_regions_optical} show the $B_{\rm PAH}$ measurements for all 1.5~kpc regions in every galaxy that have at least 25 pixels that are PAH-dominated compared to $\log_{10}($sSFR$)$ and $\log_{10}$([NII]/H$\alpha$) in gray with a function fit to the binned \refedit{medians of the} data points. Regions with 25 data points can reach a maximum signal-to-noise value of 5 under Poisson statistics, where the error scales as $\sqrt{\text{N}}$. For this reason, we exclude any regions with fewer than 25 points to ensure that the slope measurement is not dominated by noise. Even though the same cuts are made to identify PAH-dominated pixels in the 1.5~kpc regions, they have more variation in their slope values than the galaxy-wide measured $B_{\rm PAH}$ values due to lower numbers of points. In regions with low sSFR values, the slope value is very similar to $B_{S23}=1.6$ measured in \citet{Sandstrom2023}, but high sSFR regions can have slope values of up to 2 and above. The $B_{\rm{PAH}}$ values have a Spearman rank correlation coefficient of 0.38 and $-$0.30 for sSFR and the optical line ratio, respectively, at high statistical significance ($p\ll0.03$). 

The fit to sSFR can be described by the Equation 
\begin{equation}
    B_{\text{PAH}} = 0.17 \pm 0.01 \times \log_{10}(\mbox{sSFR} / {\rm yr^{-1}})+ [3.5 \pm 0.1]
    \label{eq_ssfr}
\end{equation}
and the [NII]/H$\alpha$ Equation is 
\begin{equation}
    B_{\text{PAH}} = -0.54 \pm 0.06 \times \log_{10}(\mbox{[NII]/H$\alpha$})+ [1.6 \pm 0.02].
    \label{eq_nii}
\end{equation}
The fit parameters and errors were calculated on binned medians of the data using \texttt{lmfit} \citep{Newville2014}, a non-linear least-squares minimization fitting function. In Figure~\ref{fig:megatable_regions}, the filled points in various shapes and colors show the average slope and sSFR values from all of the regions in each galaxy. The results of treating each galaxy as one large aperture, as described in Section \ref{sec:method} are plotted in the same colors and shapes with open faces. While some of the galaxy-wide and region-averaged measurements are the same, we see slight variations in some galaxies from weighting each region equally in this measurement, regardless of the number of points. Figure \ref{fig:megatable_regions_optical} shows only the averages of the optical line ratio in all the regions of each galaxy in the colored points \refedit{because these values are not included in the 1.5~kpc regions}. \refedit{This shows a significant correlation between the ratio of the PAH-correlated emission in F360M relative to F335M \refedit{(i.e. B$_{\rm{PAH}}$) with} sSFR and [NII]/H$\alpha$, likely driven by changes in radiation field hardness.} We \refedit{further} interpret the cause of the slope variations in Section \ref{sec:ssfr}.

\input{table_correlations_spear}
\subsection{Creating F335M$_{\rm PAH}$ Maps With a Variable $B_{\rm PAH}$}
  \begin{figure*}[ht]
      \centering
      \includegraphics[width = \textwidth]{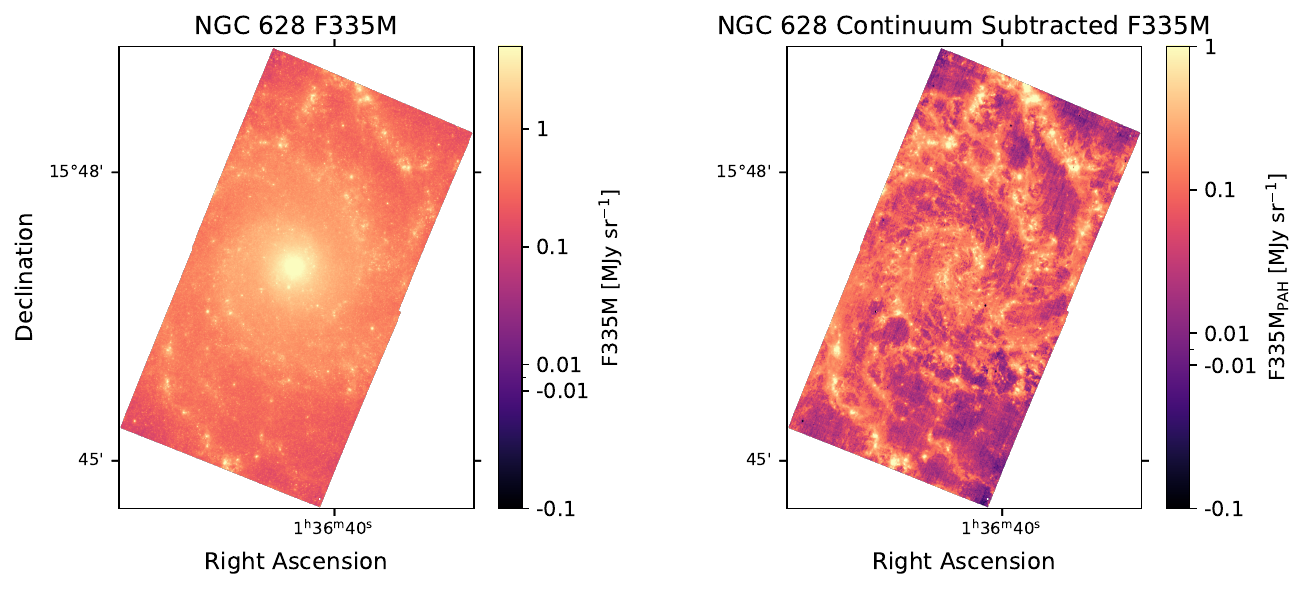}
      \caption{Left: F335M image of NGC\,628. Right: F335M$_{\rm PAH}$ map for NGC\,628 made using the slope determined for the full map of NGC\,628 on the same color scale. All 19 F335M$_{\rm PAH}$ and continuum maps are shown in Appendix Section \ref{sec:maps}.}
      \label{fig:ngc4321_ex}
  \end{figure*}
Having identified that the $B_{\rm PAH}$ slope is a function of environment, we now consider how to create continuum-subtracted F335M maps where the slope used to determine the decontaminated continuum measurement depends on position. \refedit{We created two versions of the F335M$_{\rm PAH}$ maps, one where $B_{\rm{PAH}}$ changes as a function of galactocentric radius, with each radial bin having its own calculated $B_{\rm{PAH}}$ value based on the sSFR within that annulus, and one where $B_{\rm{PAH}}$ is constant for the entire galaxy, where $B_{\rm{PAH}}$ was calculated by treating the entire galaxy as a single aperture presented in Table \ref{tab:table}.}
In F335M$_{\rm{PAH}}$ detected pixels, we find a maximum percent difference of up to 21\% in places with high sSFR values between these mapping methods, with average percent differences between maps ranging between 2--5\% in each galaxy. The result of using a galaxy-wide measured slope to create F335M$_{\rm PAH}$ maps for an example galaxy, NGC\,628, is shown in Figure~\ref{fig:ngc4321_ex}. The F335M$_{\rm PAH}$ maps and corresponding continuum maps for all galaxies in the sample are shown in Appendix~\ref{sec:maps}. The remainder of this analysis, except for the F335M$_{\rm PAH}$ maps displayed in Figure \ref{fig:ngc4321_ex} and in Appendix~\ref{sec:maps}, is done with the annular region version.

We find small differences in the maps created with the method presented here when compared to \citet{Sandstrom2023}. Across the entire sample of galaxies, the mean of the percent differences is 6\% in places \refedit{above the cuts described in Section \ref{sec:processing}.}
Individual galaxy values range between 0--16\%, with our method correcting for the changing value of PAH contamination in the F360M filter. 

When compared to \citet{Lai2020}, we find similar differences in our maps as those found in Figure 5 of \citet{Sandstrom2023}. Our maps include \refedit{PAH-correlated emission in the F335M filter, which is removed by the \citet{Lai2020} method.}
We find a much larger average percent difference across the parts of the maps containing PAH emission of 46\%, with individual galaxy percent difference values between 38--66\%. 

For users without maps of sSFR, we also define PAH slope equations for continuum subtraction based on WISE filters as sSFR proxies. WISE1 traces stellar mass while WISE4 traces star formation, the ratio of the two have been used as a measure of sSFR \citep{Leroy2019}. The WISE4/WISE1 filter ratio is significantly correlated with $B_{\rm PAH}$, with more scatter in the fit equation. The WISE filter-based Equation is 
  \begin{equation}
     B_{\rm PAH} = 0.07 \pm 0.04 \log_{10} (\text{WISE4/WISE1}) + [1.7 \pm 0.03].
 \end{equation}

The sSFR and [NII]/H$\alpha$-based Equations have R-squared values of 0.94, compared to measured slope value, while the WISE filters give an R-squared value of 0.79. The WISE filters are not perfect proxies for sSFR, so there is considerable scatter, causing the fit line to have lower R-squared values. For this reason, we recommend the use of Equations \ref{eq_ssfr} and \ref{eq_nii} where possible. For observations with only F300M and F335M data available, we also describe the creation of F335M PAH maps for galaxies in the Cycle 2 PHANGS Treasury \citep[GO 3707: PI Leroy,][]{Chown2025a} with only those filters in Appendix \ref{sec:without_360}.
 
\section{PAH Band Ratio Variations} \label{sec:results}

  \begin{figure*}[ht]
      \centering
      \includegraphics[width = \textwidth]{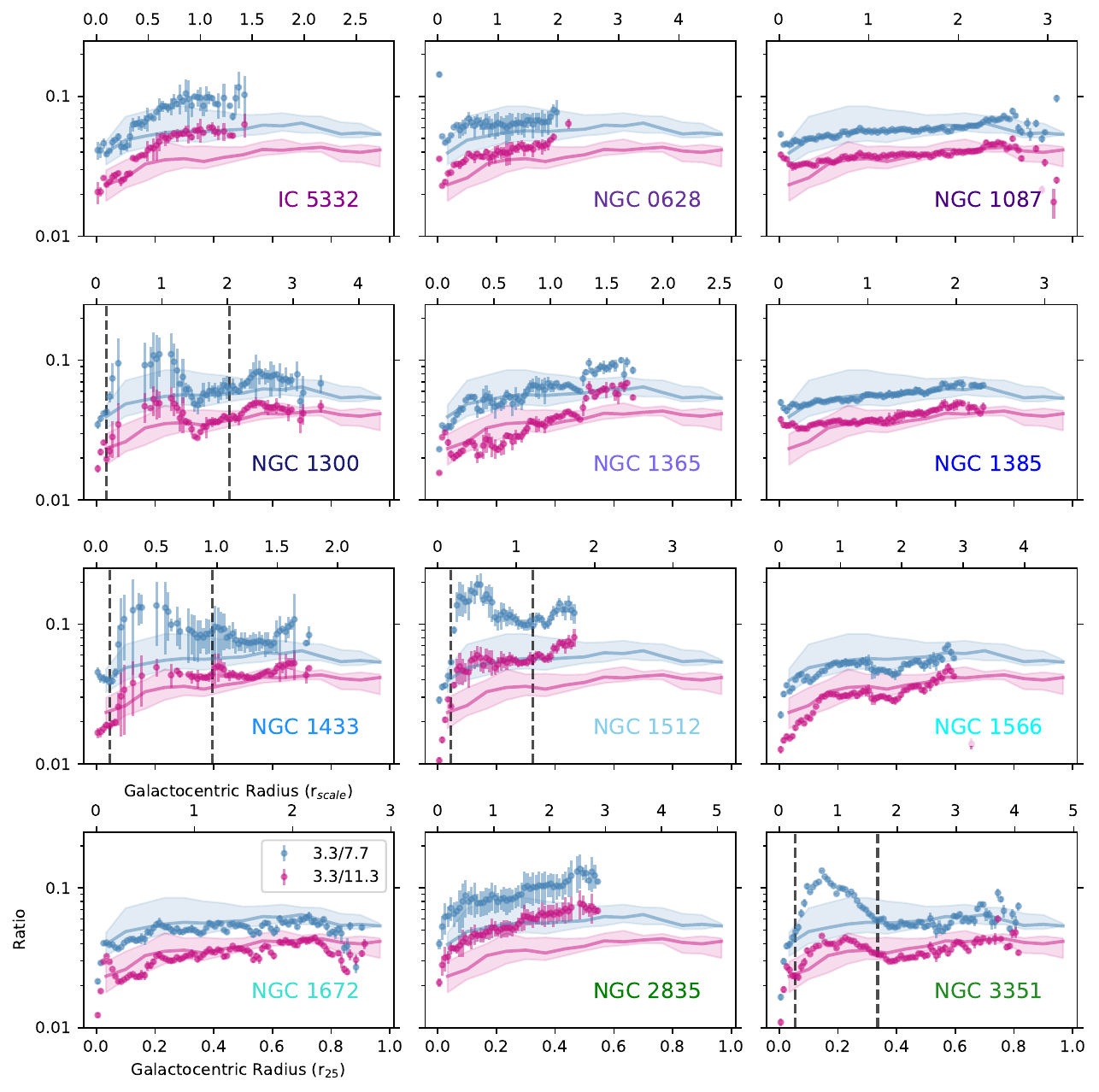}
      \caption{The ratio of the 3.3 to 11.3~$\micron$ (pink) and 7.7~$\micron$ (blue) PAH feature fluxes in each galaxy as a function of deprojected galactocentric radius measured in $r_{25}$ (bottom) and scale radius (top) units. Black vertical dashed lines represent the radial extent of ``star formation deserts''\refedit{, regions of very low SFR associated with bars,} in each galaxy that has one from \citet{Pathak2026}. \refedit{We include points above the cuts described in Section \ref{sec:processing}.}
      Bins are 0.01~$r_{25}$ in size, and error bars were calculated using the difference in values using the different mapping methods described above added in quadrature to the standard deviation divided by the $\sqrt{\rm{N}}$ in each bin. The solid line shows the trend of the ratios with galacocentric radius across all 19 galaxies in the sample, with the shaded regions surrounding them showing the 25th-75th percentiles.}
      \label{fig:ratio_radius}
  \end{figure*}
    
    \begin{figure*}[ht]
      \centering
      \includegraphics[width = \textwidth]{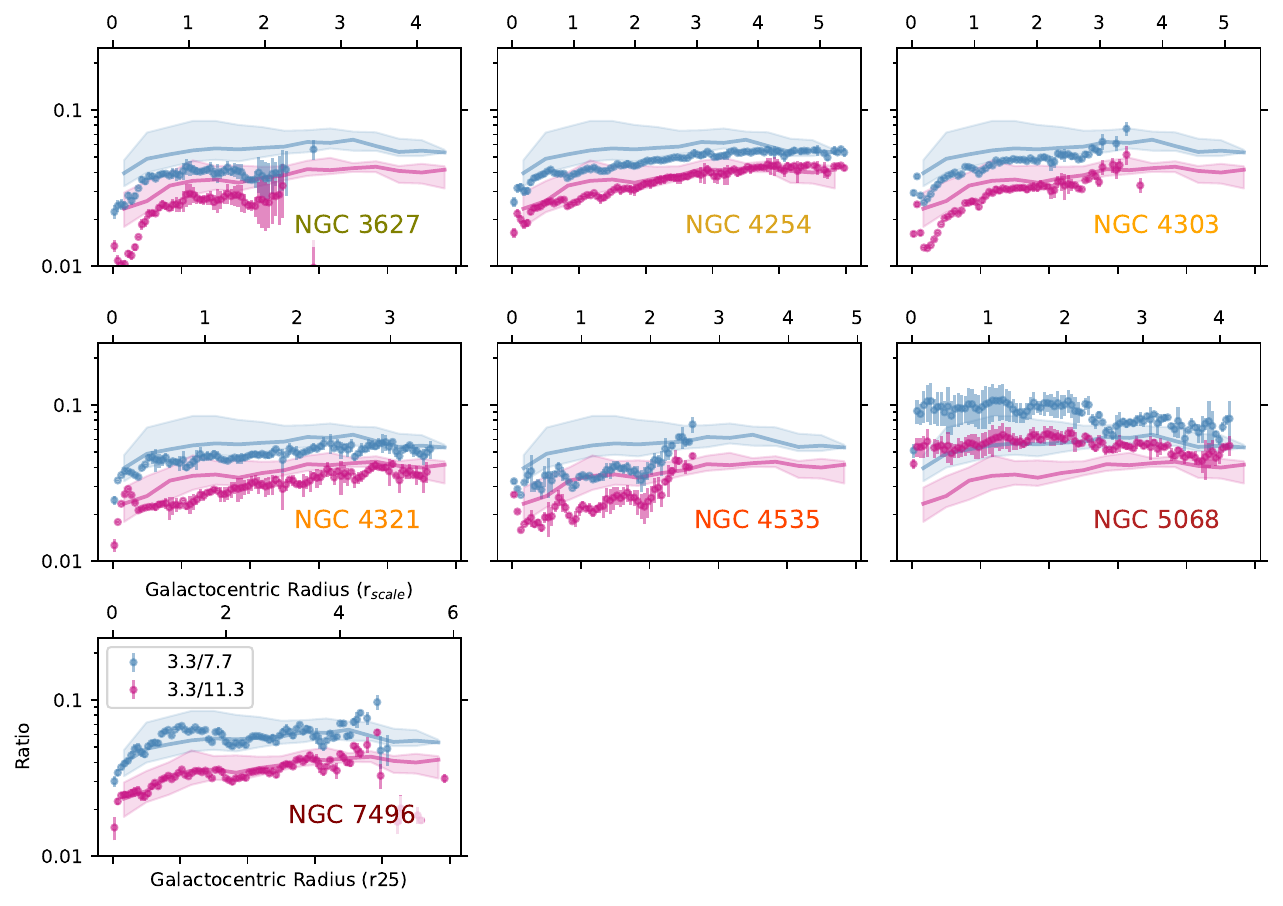}
      \caption{Figure \ref{fig:ratio_radius} (Continued).}
      \label{fig:ratio_radius2}
  \end{figure*}

To investigate the 3.3 (F335M$_{\rm PAH}$) to 7.7 (F770W$_{\rm ss}$) and 11.3~$\micron$ (F1130W) ratio values across each galaxy, we compare to several different environmental properties. We map the ratios as a function of galactocentric radius, compare with several properties on 1.5~kpc scales, bin by metallicity values using the gradient maps presented by \citet{Williams2022}, and compare to several optical line ratios at 150~pc scales. In this section we investigate F335M$_{\rm{PAH}}$/F770W$_{\rm{ss}}$ (3.3/7.7) and F335M$_{\rm{PAH}}$/F1130W (3.3/11.3) \refedit{ratios, in f$_\nu$ units (i.e. MJy~sr$^{-1}$)}, across the diverse environments included in the full PHANGS-JWST Cycle 1 sample. \refedit{While we refer to these ratios as 3.3/7.7 and 3.3/11.3~$\micron$, in this work we use all PAH-correlated emission in F335M, the stellar-subtracted F770W, and all of F1130W, so these will not be an exact match to spectral decomposition.}

\subsection{Environmental Dependence} \label{sec:environmental_dependence}
To compare the PAH band ratio values to their locations within each galaxy, we bin the PAH maps into 0.01~$r_{25}$ bins and take the ratio of the binned values. We include points \refedit{above the cuts described in Section \ref{sec:processing}.}
We use the measured differences between the annular and galaxy-wide F335M$_{\rm PAH}$ map versions added in quadrature to the standard deviation divided by the $\sqrt{\rm{N}}$ \refedit{(the standard error of the mean)} in each bin as constraints on the errors for the radial profiles. Figures \ref{fig:ratio_radius} and \ref{fig:ratio_radius2} show the 3.3~$\micron$ to 7.7 and 11.3~$\micron$ feature ratios across all deprojected galaxies when radially binned at the F1130W resolution. The solid line traces the PAH band ratio trends across the entire sample of 19 galaxies using binned medians, with the 25th-75th percentiles shown in shaded regions around the lines, showing a shallow increase of both ratios with galactocentric radius. We measure a median 3.3/11.3~$\micron$ ratio value of 0.038 and standard deviation 0.012. The 3.3/7.7~$\micron$ ratio is higher, with a median value 0.058 and standard deviation 0.025 across all galaxies in the sample. 

\refedit{We observe two main deviations away from the galaxy-wide average PAH band ratio behaviors. Elevated band ratios are seen in galaxies with ``star formation deserts'' and in three galaxies without them.}
NGC\,1300, NGC\,1433, NGC\,1512, and NGC\,3351 have been identified as having ``star formation deserts'' in \citet{Baron2025}; see also \citet{Pathak2026}. These are regions of galaxies occupied by bars with very little H$\alpha$ emission and strongly suppressed star formation rates with old stellar populations \citep{James2015}. \citet{Baron2025} found that these regions had \refedit{anomalous} behavior in the 7.7/11.3 ratio, and therefore, we investigate them here with additional scrutiny. These areas are shown in Figure \ref{fig:ratio_radius} between the vertical dashed black lines. In these regions, we see a 3.3/7.7~$\micron$ ratio much higher than the average of all the galaxies. We also see higher 3.3/11.3 and 3.3/7.7~$\micron$ ratios in IC\,5332, NGC\,2835, and NGC\,5068, spanning across a wider range of radial values. We discuss possible causes for these elevated band ratios in Section \ref{sec:discussion}.
  
\input{table_correlations_band_ratios_spear}
We compare the PAH band ratios with various environmental properties on 1.5~kpc scales shown in Table \ref{tab:band_ratios} and find strong trends with several. \refedit{B}oth ratios have a correlation with metallicity, with a significant correlation between increasing metallicity values and decreasing 3.3~$\micron$ feature strength relative to both 7.7 and 11.3~$\micron$. We see no correlation with $\Sigma_{\rm mol}/\Sigma_{\rm gas}$ for both 3.3/11.3 and 3.3/7.7~$\micron$ ratios. We also test the correlation of three commonly measured optical line ratios, [OIII]/H$\beta$, [NII]/H$\alpha$, and [SII]/H$\alpha$. The [NII]/H$\alpha$ ratio shows a significant correlation with the 3.3/11.3~$\micron$ ratio, while the [SII]/H$\alpha$ ratio shows a correlation with 3.3/7.7~$\micron$ ratio at high significance. \refedit{Only the 3.3/7.7~$\micron$ ratio has a statistically significant correlation with sSFR.}
 
 \begin{figure*}[ht]
      \centering
      \includegraphics[width = \textwidth]{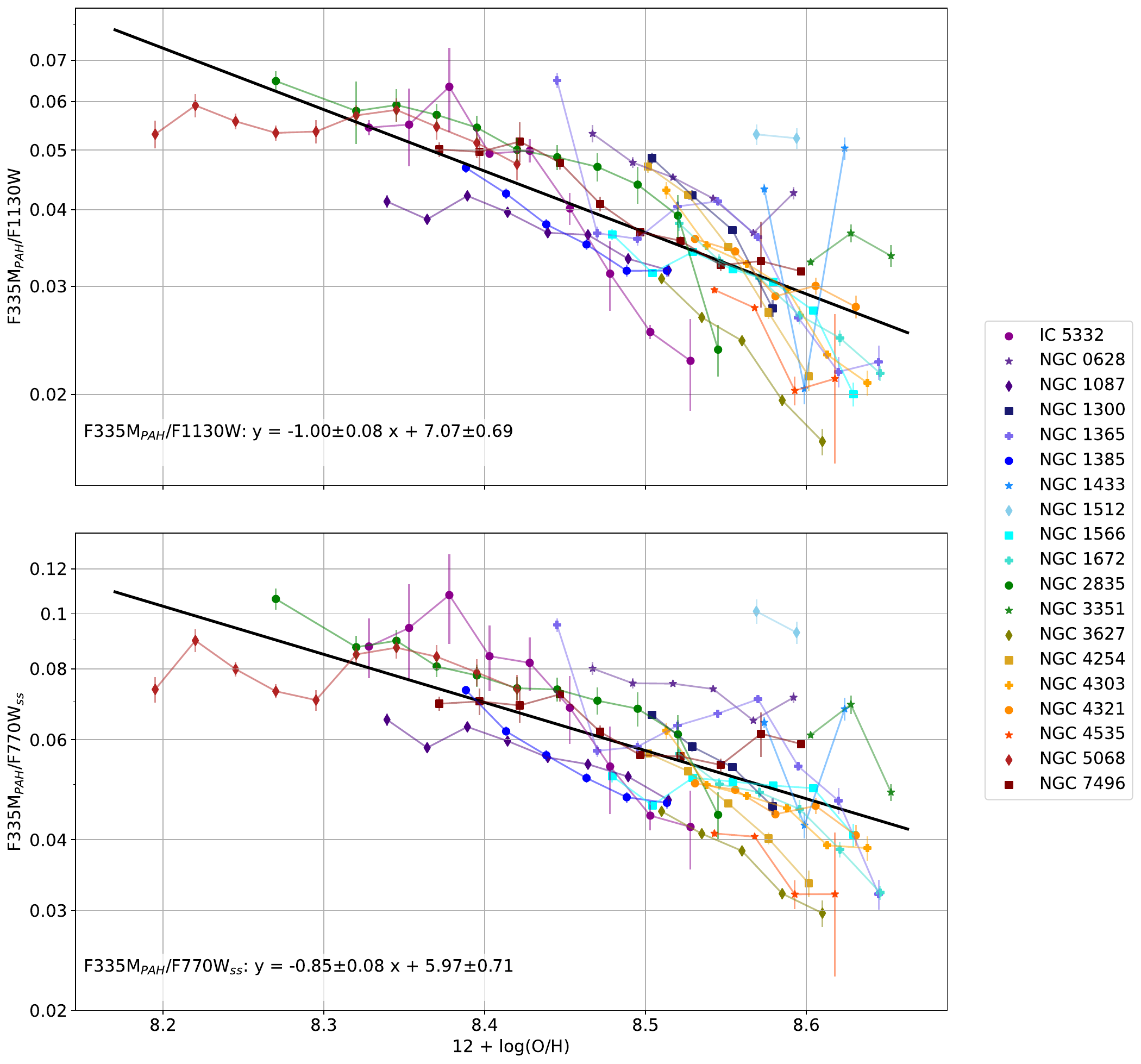}
      \caption{F335M$_{\rm{PAH}}$/F1130W (top) and F335M$_{\rm{PAH}}$/F770W$_{\rm{ss}}$ (bottom), in bins of gas phase metallicity for each galaxy at F1130W resolution. Points show the ratio of the medians calculated from points \refedit{passing the cuts described in Section \ref{sec:processing}}. Data are binned by metallicity with bins 0.025 dex in size. A trend of decreasing relative F335M$_{\rm{PAH}}$ relative to the F770W$_{\rm{ss}}$ and F1130W with increasing metallicity can be seen \refedit{across the entire sample}, which is discussed in Section \ref{sec:size}. \refedit{Variations in this relation can be seen on the individual galaxy level, with some galaxies showing monotonic decreases in relative 3.3~$\micron$ emission with increasing metallicity while others do not.} Error bars from the difference in binned values when using different F335M$_{\rm PAH}$ maps from galaxy-wide and annulus-wide sSFR values are added in quadrature to the standard deviation in each bin divided by $\sqrt{\rm{N}}$ in the bin. The data are fit with a line shown in black with the fit parameters displayed on each panel.}
      \label{fig:metallicity}
  \end{figure*}

To investigate the strong PAH band ratio correlations with metallicity, we bin \refedit{the maps} by metallicity from the \citet{Williams2022} maps in Figure \ref{fig:metallicity}. This reveals a global trend across all 19 galaxies in our sample of decreasing 3.3~$\micron$ strength relative to 7.7 and 11.3~$\micron$ with increasing values of metallicity. The 3.3/11.3~$\micron$ trend can be described by the line $y = -1.00 \pm 0.08x+7.07\pm 0.69$ and the 3.3/7.7~$\micron$ trend is best fit by $y=-0.85\pm 0.08x + 5.97 \pm 0.71$ where $x$ is metallicity in units of 12 + log(O/H). \refedit{While the trend of increasing 7.7 and 11.3~$\micron$ emission relative to the 3.3~$\micron$ feature with increasing metallicity is observed across the entire sample, this relationship is not uniform within each galaxy.}
Several galaxies, including IC\,5332, NGC\,2835, and NGC\,5068, extend to low metallicity values, allowing us to see the shorter wavelength PAH features increase in relative strength. NGC\,2835 covers 3.3/11.3~$\micron$ values from $\sim$0.065 down to 0.025 and 3.3/7.7~$\micron$ values of $\sim$0.11 to 0.05 in a single galaxy. \refedit{Given the observed metallicity dependence on PAH band ratios, the galaxies that reach the lowest values in metallicity deviate from the galaxy-average ratio trends.
Out of these three galaxies, both IC\,5332 and NGC\,5068 show non-monotonic decreases of relative 3.3~$\micron$ feature strength compared to the other PAH features.}

    \begin{figure*}[ht]
      \centering
      \includegraphics[width = \textwidth]{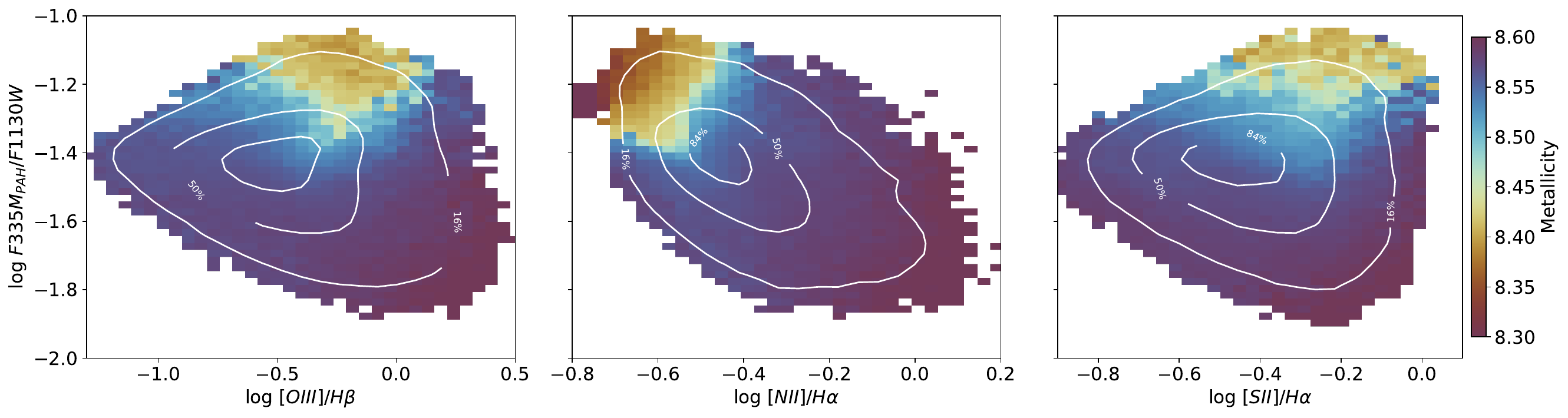}
      \caption{The 3.3/11.3~$\micron$ ratio plotted against three optical line tracers, [OIII]/H$\beta$, [NII]/H$\alpha$, and [SII]/H$\alpha$. The 2D histogram shows metallicity, with contours representing the 16th, 50th, and 84th percentiles of the counts of data shown in black. All maps are put at 150~pc resolution, \refedit{with pixels included above the cuts described in Section \ref{sec:processing}.} In the high metallicity regions we see \refedit{a correlation}
      between radiation field hardness traced by \refedit{[NII]/H$\alpha$} and F335M$_{\rm{PAH}}$/F1130W. Lower metallicity regions skew this trend most prominently when measured by [OIII]/H$\beta$ and [SII]/H$\alpha$.}
      \label{fig:optical_lines}
  \end{figure*}
  \begin{figure*}[ht]
      \centering
      \includegraphics[width = \textwidth]{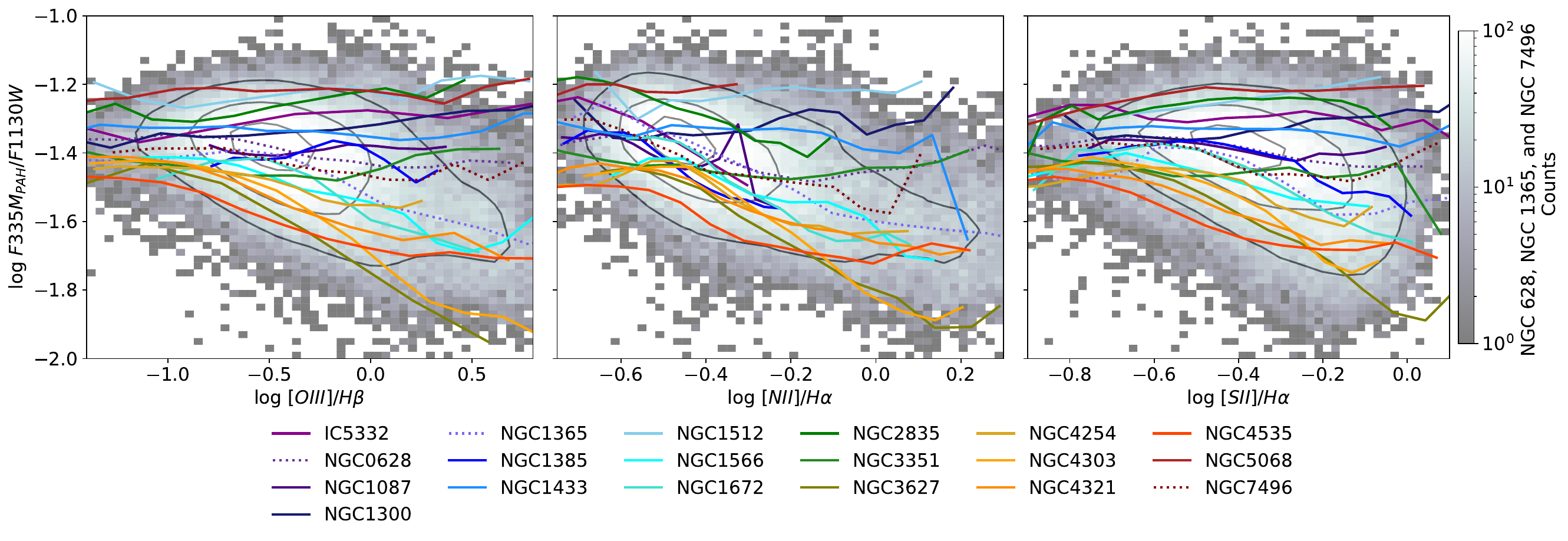}
      \caption{The 3.3/11.3~$\micron$ ratio plotted against the same optical line traces as Figure \ref{fig:optical_lines} to compare to \citet{Baron2024}. Binned median individual galaxy trends in these optical line ratios are each represented with their own color. The galaxies with the dotted lines are shown as the background histogram and contours. These are the three galaxies used in the \citet{Baron2024} analysis, NGC\,628, NGC\,1365, and NGC\,7496. The contours represent the 16th, 50th, and 84th percentiles of the background histogram of the three galaxies in \citet{Baron2024}. All maps are put at 150~pc resolution, \refedit{and pass the cuts described in Section \ref{sec:processing}.} The 2D histogram shows good agreement with the analysis done by \citet{Baron2024}. When the analysis is expanded to include the full sample of 19 galaxies, we observe a flattening of the trends discussed in Section~\ref{sec:environmental_dependence}.}
      \label{fig:optical_lines_individual}
  \end{figure*}

    \begin{figure*}[ht]
      \centering
      \includegraphics[width = \textwidth]{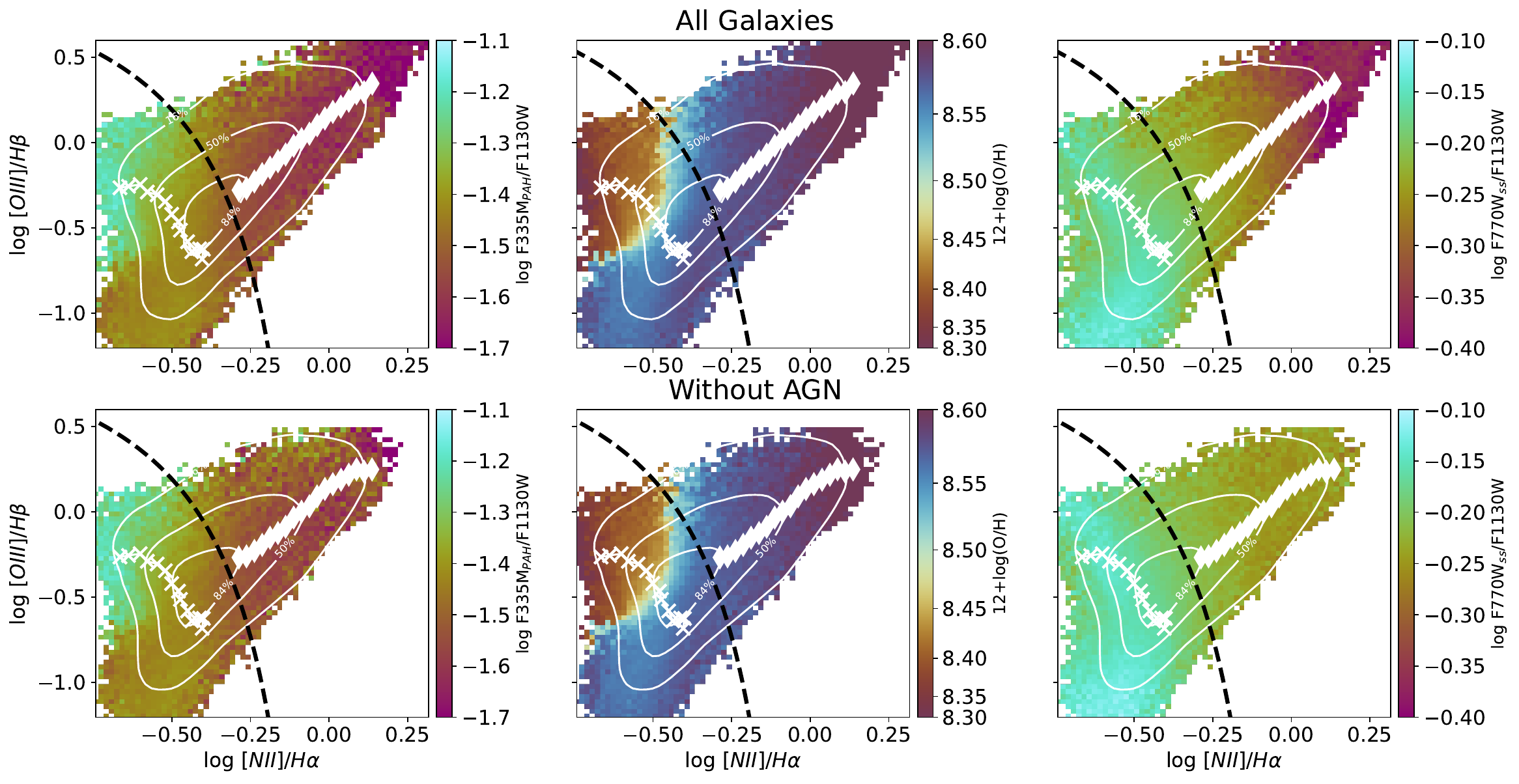}
      \caption{BPT \citep{baldwin1981}, diagram created using the 150~pc data \refedit{color coded by} three quantities \refedit{with black contours showing where the 16th, 50th, and 84th percentiles of the data lie. Top panels include data from all galaxies and the bottom panels exclude galaxies that have been identified as AGN hosts, NGC\,1365, NGC\,1672, NGC\,4303, and NGC\,7496.} The color bar on the left panel shows the band ratio 3.3/11.3~$\micron$, the middle panel is colored by metallicity to highlight where the lowest metallicity regions fall in the diagram\refedit{, and t}he right panel shows the 11.3/7.7~$\micron$ ratio. The \refedit{left of the} black dashed line on all panels represents where star-forming \refedit{regions} fall on the diagram \citep{Kauffmann2003}. We show two distinct tracks in PAH band ratio trends in the star-forming, left, and LINER, right, divided by the  \citet{Kauffmann2003} line. In star-forming regions, we bin points by metallicity, plotted as `x' marks, and calculate the median values of both PAH band ratios and optical line ratios in each bin. In the LINER regime, we do a similar analysis, binned by log([NII]/H$\alpha$). We observe a strong PAH band ratio dependence on metallicity for bins in the star-forming regime while those in the LINER regime are more heavily influenced by radiation field hardness.}
      \label{fig:bpt}
  \end{figure*} 

\begin{figure*}[ht]
      \centering
      \includegraphics[width = \textwidth]{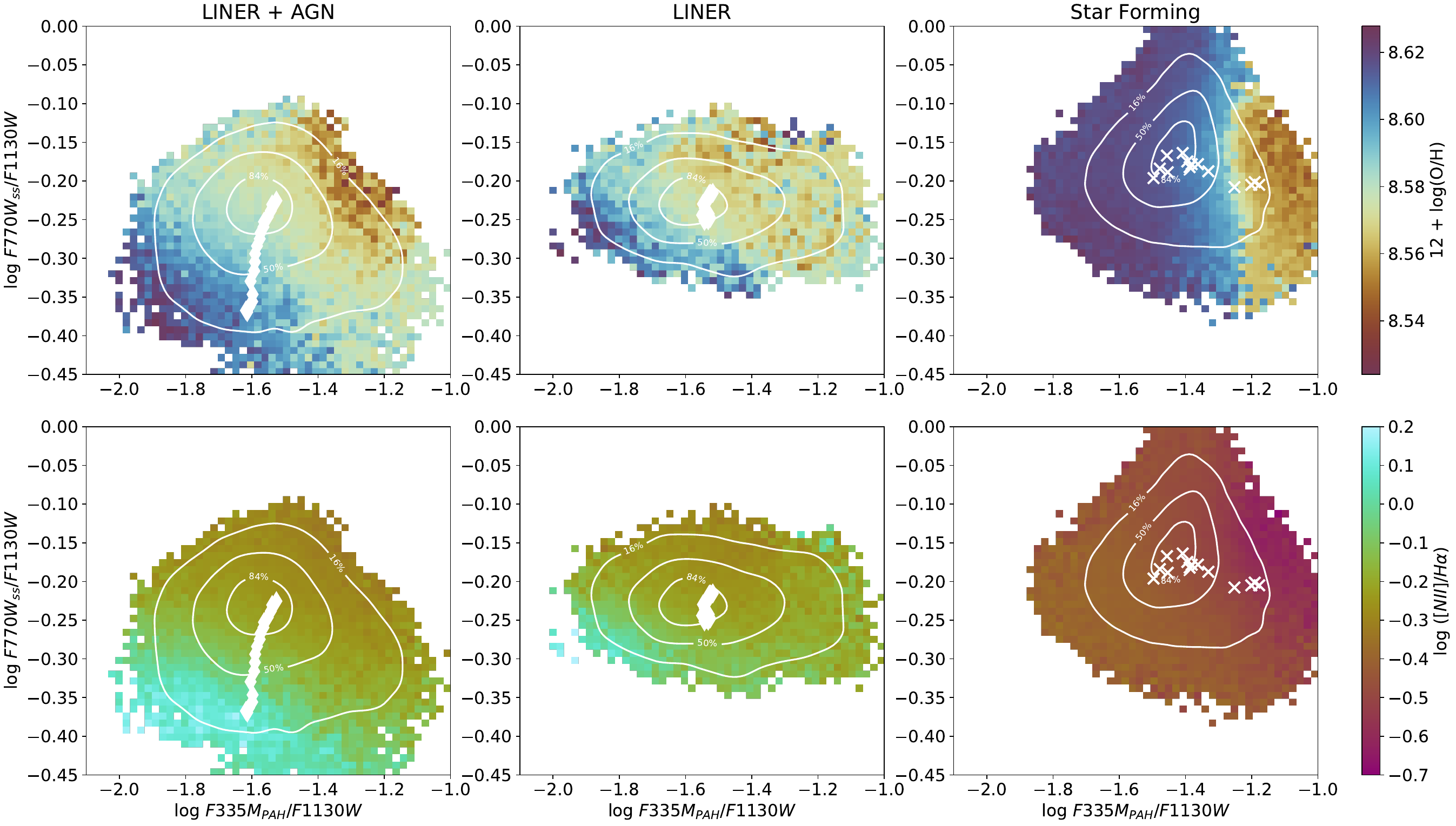}
      \caption{7.7/11.3 plotted against 3.3/11.3 at 150~pc resolution colored by metallicity (top) and [NII]/H$\alpha$ (bottom) for comparison to \citet{Baron2024}. We isolate the points that are considered LINERs \refedit{and AGN (left), LINERs (middle),} and star-forming (right). Diamond points are plotted in the left panels represent binned medians 0.025 width in log([NII]/H$\alpha$). `x' points in the right panels show binned medians in metallicity. Contours represent the 16th, 50th, and 84th percentiles. The \refedit{LINER$+$AGN} region points show a trend between 7.7/11.3~$\micron$ band ratios with [NII]/H$\alpha$ \refedit{and metallicity}, while the 3.3/11.3~$\micron$ ratio remains relatively constant. In the star-forming regime, the 3.3/11.3~$\micron$ band ratio is correlated with metallicity and the 7.7/11.3~$\micron$ \refedit{ratio shows little change across these points.}}
      \label{fig:band_ratios}
  \end{figure*} 
  
To investigate the correlation between the PAH band ratios and radiation field spectrum, we compare the 3.3/11.3~$\micron$ ratio against [OIII]/H$\beta$, [NII]/H$\alpha$, and [SII]/H$\alpha$ on 150~pc scales, shown in Figure \ref{fig:optical_lines}, color coded by metallicity. We see the strongest correlation with [NII]/H$\alpha$, with a Spearman correlation value of $-$0.36 and $p$-value $\ll$0.03. 

\citet{Baron2024} observed correlations between the 3.3/11.3~$\micron$ ratio with optical line ratios in three PHANGS galaxies, NGC\,628, NGC\,1365, and NGC\,7496. They found that the correlations with [OIII]/H$\beta$ and [NII]/H$\alpha$ had more scatter than the tight correlation with [SII]/H$\alpha$. In this work, the [OIII]/H$\beta$ and [SII]/H$\alpha$ correlations are skewed from those observed in \citet{Baron2024} by a group of points falling at a lower metallicity. For direct comparison with \citet{Baron2024}, Figure \ref{fig:optical_lines_individual} shows the same three galaxies included in their study, NGC\,628, NGC\,1365, and NGC\,7496, in the background histogram. Binned median averages of each individual galaxy are included in this sample are plotted over in different solid colored lines, with the three galaxies repeated in this analysis in dotted lines. With our new prescription for F335M continuum subtraction, we still see very similar trends in the three galaxies included in their work, meaning our method is not driving the observed changes in the trends when compared to their work. The diverse range of environments covered in the complete sample of 19 galaxies, especially those with raised 3.3/11.3~$\micron$ ratios such as IC\,5332, NGC\,2835, and NGC\,5068, flattens out the trends in [OIII]/H$\beta$ and [SII]/H$\alpha$. These three galaxies \refedit{reach the lowest} gas-phase metallicity values, causing variation in PAH band ratio values. 

To investigate the dependence of band ratio values on radiation field hardness, Figure \ref{fig:bpt} shows the diagnostic BPT diagram \citep{baldwin1981} color coded by metallicity (center) and 3.3/11.3 (left) and 7.7/11.3~$\micron$ (right) PAH band ratios. \refedit{We show two versions of this plot, one including all galaxies in the top three panels, and one where we remove galaxies that have been identified as AGN hosts in the bottom three panels; these galaxies include NGC\,1365, NGC\,1672, NGC\,4303, and NGC\,7496.}
This combination of optical lines have been shown to differentiate the source of the ionizing radiation between star-forming HII regions and older stellar population dominated, low-ionization nuclear emission-line regions (LINERs), \refedit{and AGN.} \refedit{We plot } a line from \citet{Kauffmann2003} approximately dividing the \refedit{star forming and LINER} regions. \refedit{Here we refer to the regions identified by BPT diagrams after removing AGN, typically called LINERs. \citet{Belfiore2022} studied this component extensively and found that it originates from diffuse ionized gas distributed across the galaxies on kpc scales, which they suggest is powered by leaking ionizing radiation from HII regions and ionization from hot and evolved stellar populations.}

Examining these color-coded BPT figures, we see two main axes that have a\refedit{n} effect on the PAH band ratios. The first traces a metallicity sequence in regions dominated by ionization by young, massive stars \citep[i.e. to left of the line from][]{Kauffmann2003}. The second sequence traces the hardness of the radiation field as we move from photoionization dominated by HII regions into the LINER \refedit{and AGN} regime towards the right side of the \citet{Kauffmann2003} line. Two tracks are plotted in the two regimes of the BPT diagram in Figure \ref{fig:bpt}. Points plotted as `x' marks are binned in metallicity in the star forming regime and diamond points are binned by log([NII]/H$\alpha$) in the LINER regime \refedit{with and without AGN. M}edian values of all other quanitites including PAH band ratios, metallicity, and optical line ratios are calculated within each bin. We see a spread in metallicity on the left side of the \citet{Kauffmann2003} line, while the right side remains at a constant, high value of metallicity but shows a spread in radiation field hardness. In regions consistent with being illuminated by HII regions, we find a strong trend of 3.3/11.3~$\micron$ PAH band ratios, with increasing ratios for decreasing metallicities. On the other side of the \citet{Kauffmann2003} line, where regions are illuminated by older stellar populations, the 3.3/11.3~$\micron$ band ratio values are lower and do not appear to have a clear metallicity dependence. The 7.7/11.3~$\micron$ trend is primarily along a diagonal direction in the BPT space, following the transition from HII region to LINER \refedit{and AGN} dominated emission, with a less clear dependence on metallicity. 

To visualize \refedit{the PAH band ratio's dependence on metallicity and radiation field spectrum more clearly,} in Figure \ref{fig:band_ratios} we show two PAH band ratios, 7.7/11.3 and 3.3/11.3~$\micron$, colored by metallicity and [NII]/H$\alpha$ for regions identified by the BPT diagram as star forming\refedit{, LINERs and AGN, and just LINERs}. In the star forming regions the 3.3/11.3~$\micron$ ratio varies primarily horizontally in this space, correlated with metallicity. \refedit{In the LINER and AGN regime, optical line ratios show a metallicity dependence, and correlated PAH band ratio change. To separate the effects of both radiation field spectrum and metallicity, binning in BPT space is necessary, such as in Figure \ref{fig:bpt}. In galaxies with little metallicity variation, PAH band ratios should, in principle, provide information on the spectral shape of the radiation field.}
  
\subsection{Model Comparison to \citet{Draine2021}} \label{sec:models}
  \begin{figure*}[ht]
      \centering
      \includegraphics[width = 0.85\textwidth]{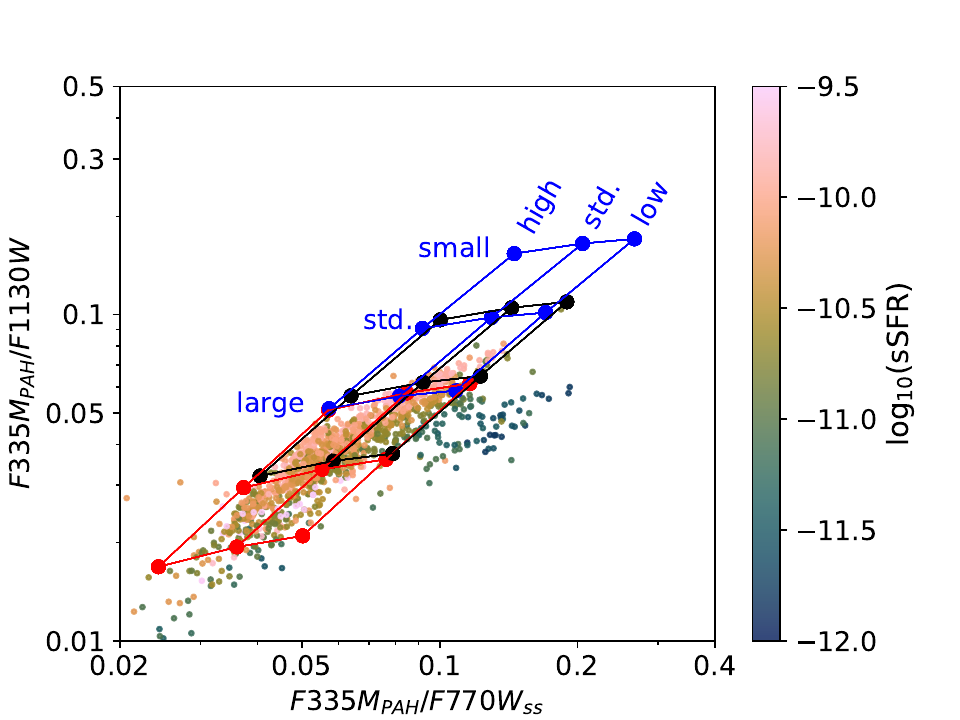}
      \caption{The ratio of the 3.3/7.7~$\micron$ PAH features versus the ratio of the 3.3/11.3~$\micron$ PAH ratios at the F1130W resolution colored by the average sSFR in each bin. Each point is binned radially in the same process as Figures \ref{fig:ratio_radius} and \ref{fig:ratio_radius2} with bin sizes 0.01~$r_{25}$. Dust model grids from varying radiation fields, including the M31 bulge \citep{Groves2012}, mMMP \citep{mathis1983}, and BC03 3~Myr old starburst \citep{Bruzual2003} from \citet{Draine2021}, are shown in red, black, and blue, respectively. Small, standard, and large labels refer to three PAH size populations modeled in \citet{Draine2021}, while high, standard, and low label the three ionization fractions modeled. The majority of our data lies within the models grids, with the lowest sSFR points falling outside. We discuss possible reasons for this in Section \ref{sec:models}.}
      \label{fig:models_all}
  \end{figure*} 

To interpret the PAH band ratios and their dependence on PAH population size, charge, and the spectrum of the illuminating radiation field, we compare to models presented by \citet{Draine2021}. We use models with three different stellar radiation fields, including a 3~Myr starburst presented by \citet{Bruzual2003}, the solar-neighborhood radiation field typical of the diffuse ISM from \citet{mathis1983}, and starlight from an older population from observations of the M31 bulge presented by \citet{Groves2012}. We also consider three different size distributions from \citet{Draine2021} --- small, standard, and large, classified by a$_{01}$, the average radius describing the log-normal size distribution of the smallest PAHs = 3.0, 4.0, and 5.0~\AA. We also consider three different ionization levels of the PAH population: the standard ionization refers to an ionization fraction, $f_{\rm ion}$, with a similar value to \citet{Draine2007}, with low and high $f_{\rm ion}$ referring to a factor-of-two shift in either direction.

We perform synthetic photometry on the \citet{Draine2021} models using the NIRCam and MIRI transmission filters to match our data. We use the Equation 
\begin{equation}
    F^{\text{synth.}}_{\text{filter}} = \frac{\int^{\lambda_1}_{\lambda_0}(F_\nu (\lambda)\lambda/hc)t_{\lambda, \text{filter}}d\lambda}{\int^{\lambda_1}_{\lambda_0}(\lambda/hc)t_{\lambda, \text{filter}}d_\lambda}
\end{equation}
to calculate the synthetic fluxes, $F_\nu(\lambda)$, for an imaging filter in units of MJy~sr$^{-1}$ given in \citet{Gordon2022}. This integral is calculated over the wavelength range of the imaging filter with $t_{\lambda, \text{filter}}$ being its total throughput. The \citet{Draine2021} models do not contain stellar continuum, and the F335M and F770W data we use for comparison have had the stellar continuum removed. However, it is important to note that both the models and the F770W and F1130W observations contain hot dust continuum that has not been removed. Even though we have removed hot dust continuum from F335M, we find that the 3.3~$\micron$ feature dominates the model spectra over hot dust continuum within the F335M coverage.

Figure \ref{fig:models_all} shows all of the galaxies binned using the same method as Figures \ref{fig:ratio_radius} and \ref{fig:ratio_radius2}. Bins are 0.01~$r_{25}$ in size, compared to the model grids from \citet{Draine2021}. The span of the the 3.3/11.3 and 3.3/7.7~$\micron$ ratios is in good agreement with the models. However, the majority of the points are shifted towards lower 3.3/11.3 and 3.3/7.7~$\micron$ ratios, and fall within the span of the M31 bulge radiation field models or the large size distribution range of the MMP radiation field models. It is worth noting that the galaxies in our sample are still actively forming stars and in many regions should be dominated by a younger stellar population. 

\begin{figure}[ht]
\centering
\includegraphics[width = \columnwidth]{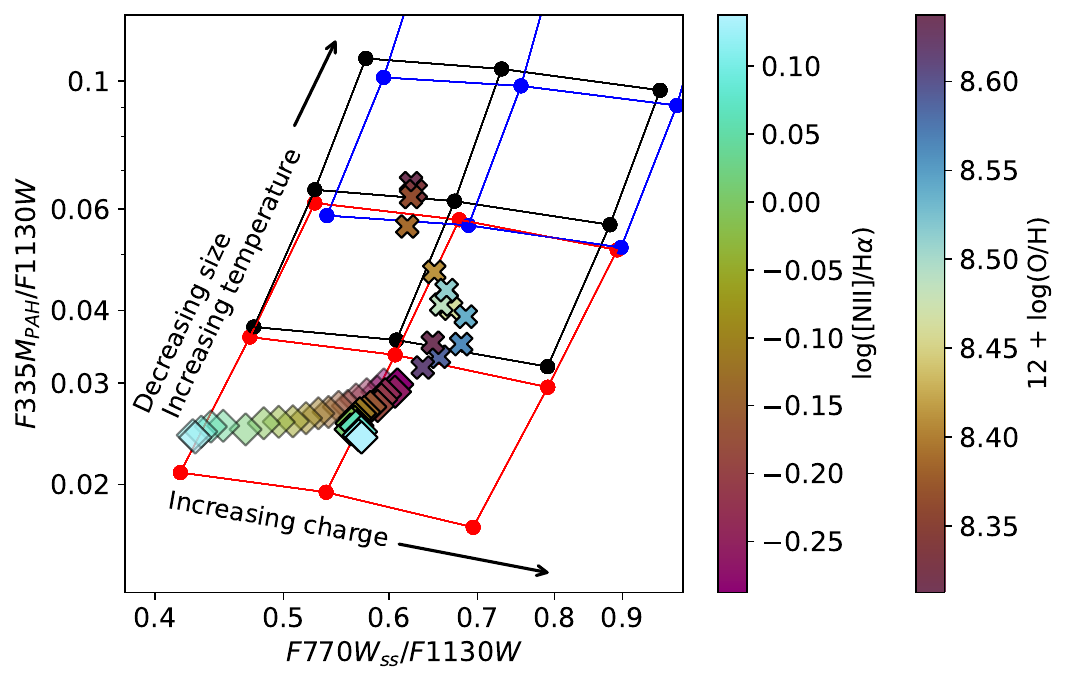}
      \caption{We plot the values of the band ratios identified by the `x' marks on Figure \ref{fig:band_ratios} colored by average metallicity in the bin on the \citet{Draine2021} model grids. 
      The diamond points are color coded and binned by [NII]/H$\alpha$ value in the LINER regions of the BPT diagram. We show two tracks of binning in [NII]/H$\alpha$. All galaxies are considered in the transparent points, while the \refedit{opaque} points exclude AGN hosts in the sample (NGC\,1365, NGC\,1672, NGC\,4303, and NGC\,7496). Directions of increasing charge, decreasing size, and increasing temperature are annotated with arrows pointing in each direction. Star-forming points binned by metallicity are consistent with a shift towards smaller PAHs as metallicity decreases. \refedit{We see the influence of the presence of AGN on the 7.7/11.3~$\micron$ ratio.} 
      After the AGN are removed, the LINER points follow a trend of increasing relative short wavelength PAH feature strength compared to the 11.3~$\micron$ feature with increasing radiation field hardness.}
\label{fig:bpt_grids}
\end{figure}

To investigate the dependence of band ratio values on size and radiation field hardness, we show the same tracks from Figures \ref{fig:bpt} and \ref{fig:band_ratios} plotted on the \citet{Draine2021} grids Figure \ref{fig:bpt_grids}. We color code the `x' points from the star forming regime of the BPT diagram by the metallicity value in each bin. The LINER points are displayed as transparent diamonds and are color coded by the binned [NII]/H$\alpha$ value. The \refedit{opaque} diamonds show the same bins in [NII]/H$\alpha$ after removing the four galaxies with AGN, NGC\,1365, NGC\,1672, NGC\,4303, and NGC\,7496. Here we plot the 3.3/11.3~$\micron$ ratio on the y-axis and the 7.7/11.3~$\micron$ on the x-axis. We see that over the spread of metallicities covered in the star-forming points, we cover a large range in 3.3/11.3~$\micron$ and a relatively small range in 7.7/11.3~$\micron$. The LINER points show a large variation only in 7.7/11.3~$\micron$ ratios as a function of [NII]/H$\alpha$ value when AGN are included. When we remove these galaxies from the binned points, we see a smaller spread in 7.7/11.3~$\micron$ values. We interpret these results in Section \ref{sec:radiation}. 

\section{Discussion} \label{sec:discussion}

\subsection{$B_{\rm PAH}$ Slope Dependence on Radiation Field Hardness} \label{sec:ssfr}
In Section~\ref{sec:bpah_meas}, we presented the variations in the slope of F335M/F300M and F360M/F300M colors in PAH dominated regions in our sample of 19 nearby star-forming galaxies. We found that $B_{\rm PAH}$ shows a significant correlation with sSFR and [NII]/H$\alpha$. The slope values represent changes in the relative strengths of the PAH or PAH-correlated features captured by the F335M and F360M filers. The F335M filter is centered on the 3.3~$\micron$ aromatic feature \refedit{and captures the 3.4~$\micron$ aliphatic feature, 3.47~$\micron$ plateau, and part of the PAH-correlated continuum}, while the F360M filter \refedit{covers part of the 3.4~$\micron$ aliphatic feature, 3.47~$\micron$ plateau, and PAH-correlated continuum, the origin of which is still not entirely understood. Further discussion of the 3.4~$\micron$ feature's significance in F360M is located in Appendix \ref{sec:3p4}.} A steeper slope corresponds to a higher F335M/F360M PAH ratio, while a shallower slope means a lower F335M/F360M PAH ratio.

All of the PAH features captured in the F335M and F360M filters may vary with environment. The 3.4~$\micron$ feature is commonly attributed to the C-H stretching mode of aliphatic bonds \citep{Pendleton2002}. It has also been suggested that the 3.4~$\micron$ feature could be contributed to by hot bands of the aromatic CH stretch \citep{Boersma2023}. Super-hydrogenated PAHs are thought to be the carriers of the 3.47~$\micron$ plateau emission \citep{Hammonds2015}. The PAH continuum in F360M (and F335M), though still uncertain, has been attributed to several types of PAH emission, including the CH aromatic stretching modes between 3.2--3.6 $\micron$ that are easily affected by anharmonic and resonance interactions \citep{Mackie2022}. The PAH continuum may also be related to the wings of longer wavelength PAH features \citep{Peeters2024}. Overtone bands of longer wavelength PAH features also contribute to the $\sim$3~$\micron$ region \citep{Allamandola2021}. 

Increasing sSFR and decreasing [NII]/H$\alpha$ values likely indicate softer ionizing radiation fields and harder FUV non-ionizing radiation illuminating the PAHs generated by young stellar populations \citep{Belfiore2023, Baron2024}. In this work, we see steeper $B_{\rm PAH}$ slope values at higher sSFR and lower [NII]/H$\alpha$, \refedit{meaning there is a weaker emission from the bands that contribute to F360M filter relative to the F335M filter}, where the intensity of star formation relative to older stellar populations is higher. 

\refedit{Several features contribute to the F360M, including the 3.4~$\micron$ feature, 3.47~$\micron$ plateau, and PAH-correlated continuum at $\sim$3.5~$\micron$ and beyond. When isolating the PAH-correlated emission in F360M, we remove the stellar continuum and other emission contributing to this filter.} If the F360M emission is dominated by aliphatic hydrocarbon emission, then our observed trend is consistent with several results in the literature about how aromatic-to-aliphatic ratios vary with SFR. Observations of the aromatic-to-aliphatic ratio in M82 revealed the abundance of aliphatic hydrocarbons in the halo  decreasing towards the star-forming disk \citep{Yamagishi2012}. A decrease in aliphatic hydrocarbons has also been observed in regions with intense radiation fields \citep{Pilleri2015}. The aromatic-to-aliphatic ratio has been observed to change by a factor of 2 across NGC\,7469, with an even lower ratio towards the AGN \citep{Lai2023}. \citet{Lai2020} predicted that the presence of star formation would decrease the aliphatic population, agreeing well with our results. 

However, if the F360M emission is tied more to overtone modes, the decrease of overtone strength with increasing sSFR would not be expected. The higher energy photons from active star formation excite both the 3.3~$\micron$ feature and additional overtone and continuum emission, with the latter enhancing the strength of the F360M relative to F335M \citep{Boersma2023, Chown2025}. Alternatively, much of what contributes to the PAH continuum underlying the F360M emission could arise from the wings of longer wavelength PAH features. In this case, to reproduce our observed environment-dependent slope, the 3.3~$\micron$ feature must become stronger relative to those longer wavelength features as the sSFR increases. In Table \ref{tab:band_ratios}, we see a significant correlation between decreasing values of 3.3/7.7~$\micron$ values with increasing sSFR, which disagrees with the observed trend if the the wings of longer wavelength PAH features enhanced the PAH emission in F360M relative to F335M. Without spectroscopic information, we cannot fully dissect which of these features shifts the PAH-correlated F360M/F335M color. We plan to investigate this work in the future with JWST NIRSpec observations.

\subsection{PAH Size Variation} \label{sec:size}
One of the most significant correlations we see with PAH band ratios in every galaxy is with metallicity. In Figures \ref{fig:ratio_radius} and \ref{fig:ratio_radius2}, the average radial trend for all the galaxies in our sample shows an increase of 3.3/7.7 and 3.3/11.3~$\micron$ ratios with increasing galactocentric radius. This can be interpreted as an effect of the observed metallicity gradient in the PHANGS galaxies, with metallicity values decreasing with increasing distance from the center. Several galaxies have PAH band ratio values above the average radial trend, some of which can be explained by the presence of star formation deserts (see Section~\ref{sec:radiation}). However, IC\,5332, NGC\,2835, and NGC\,5068 show higher PAH band ratio values without having a star formation desert. These three galaxies have lower metallicities and/or steeper metallicity gradients as shown in the \citet{Williams2022} metallicity maps. Binning the galaxies by metallicity (Figure~\ref{fig:metallicity}), we see a global trend among all 19 galaxies for the 3.3/11.3 and 3.3/7.7 ratios to increase with decreasing metallicity\refedit{, with individual galaxies showing deviations from this trend, highlighting the effects of other physical properties, such as radiation field hardness, on this relation on smaller scales. This agrees with \citet{Whitcomb2025}, where they also observed an increase in relative 3.3~$\micron$ feature strength with decreasing metallicity.} Lastly, separating the band ratios in BPT diagram space (Figure~\ref{fig:bpt} and ~\ref{fig:band_ratios}), we also see a clear metallicity dependence in the band ratios in places where HII regions dominate the photoionization.

Figure~\ref{fig:bpt_grids} summarizes the metallicity trend for the star-formation dominated side of the BPT diagram.  In this space, we see lower metallicity regions move primarily vertically, evolving mainly in the 3.3/11.3~$\micron$ ratio. Comparing to the \citet{Draine2021} grids, this shift can be explained with decreasing PAH size (e.g. approximately from ``standard'' to ``small'' in the MMP grid) or with increasing radiation field hardness (e.g.\ ``standard'' size from M31 bulge to MMP-like radiation field). To interpret these trends, we must first investigate whether the increased hardness of the radiation field of low metallicity young stellar populations could play a role in the band ratio changes.

To investigate what is driving the observed trend in 3.3/7.7 and 3.3/11.3~$\micron$ ratios with metallicity, we compare to the \citet{Draine2021} and several stellar population synthesis models using the Flexible Stellar Population Synthesis  \citep[FSPS; ][]{Conroy2009, Conroy2010} package to model stellar populations ranging in metallicity between $\sim$8-9 in 12+log(O/H) units to cover a comparable range observed in the PHANGS galaxies. We find that the FUV-optical slope, measured between 912-2000\AA~and 3000-4000\AA~\citep{Baron2024}, for the highest and lowest metallicity stellar populations only changes by 0.135 dex, corresponding to a 3.3/11.3~$\micron$ ratio change of $\sim$1.15\%. When compared to the interstellar radiation fields used to calculate the band ratios in the \citet{Draine2021} models, this is not a large enough change to be the driver of the 3.3/11.3~$\micron$ ratio variations we observe. This suggests that, according to these models, the band ratio variation we observe correlated with metallicity is driven by an intrinsic change in the size of the PAH population in our sample of galaxies. \citet{Whitcomb2024} also observed that the radiation field spectrum effects at differing metallicities were not large enough to explain this shift in band ratio values in M101, NGC\,628, and NGC\,2403.

Using the star-forming points identified in Figure \ref{fig:band_ratios}, plotted on the \citet{Draine2021} model grids in Figure \ref{fig:bpt_grids} we see a wide range of 3.3/11.3~$\micron$ values, increasing towards lower metallicities. The 7.7/11.3~$\micron$ ratio also evolves slightly with metallicity, increasing towards higher metallicity values \refedit{from the lowest metallicity values covered in our sample to 12 + log(O/H) $\approx$ 8.55, where it then curves slightly back, but is still increased relative to the lowest metallicity bin}. This may indicate a shift towards more neutral PAHs at lower metallicity. It also supports a shift to smaller sizes at lower metallicity because as PAH populations decrease in size, they also become more neutral \refedit{\citep{Li2001}}. \refedit{This is driven by two main factors: the photoionization rate scales as the cube of the radius of the PAH, $a^3$, and the electron capture rate scales as $a^2$ \citep{Draine2021}.} Our measured values of the 3.3/11.3~$\micron$ ratio decrease while the 7.7/11.3~$\micron$ ratio increases with metallicity, supporting our interpretation as a metallicity trend with PAH size and charge. 

Our observations of the trend for enhanced 3.3~$\micron$ PAH emission in low metallicity environments, shown in Figure \ref{fig:metallicity}, follow the already observed trend towards smaller PAHs at low metallicity. This trend was observed using longer wavelength band ratios from {\em Spitzer} spectra in both the {\em Spitzer} Infrared Nearby Galaxies Survey (SINGS) galaxies \citep{Smith2007} and the Small Magellanic Cloud \citep{Sandstrom2012}. While it has been seen that the more fragile small PAH population can be destroyed in low metallicity environments such as blue compact dwarfs \citep{Hunt2010}, destruction cannot explain a shift towards smaller PAH sizes, as it shifts the size distribution the other direction. After observing this trend using {\em Spitzer} spectroscopy across three galaxies with radial metallicity gradients, including NGC\,628, \citet{Whitcomb2024} modeled a metallicity-dependent PAH size distribution, where the average size of PAH grains decreases with metallicity and the mass of the grains declines below a threshold metallicity value of $Z = 0.63 \pm 0.03 Z_{\odot}$, that agrees well with our results of more small PAHs in lower metallicity regions. \refedit{However, we do not observe the same sharp increase of relative 3.3~$\micron$ emission at the metallicity threshold seen in \citet{Whitcomb2024}, possibly from different metallicity calibrations changing the threshold value.} JWST NIRCam observations of M101 presented by \citet{Whitcomb2025} find that the fraction of 3.3~$\micron$ feature strength compared to the total PAH luminosity increases from $\sim$1-3\% as metallicity decreases from Solar values to 40\% Solar. They also explore the inclusion of photo-destruction into the inhibited growth model and still find agreement with their observations. \citet{Tarantino2025} shows strong 3.3~$\micron$ emission compared to longer wavelength features in extremely low metallicity, 7\% Solar, galaxy Sextans A. Observations of II Zw 40, 25\% Solar metallicity, show an increased contribution of the 3.3~$\micron$ compared to total PAH emission \citep{Lai2025}.

A number of studies now agree, including this comprehensive look at 19 nearby galaxies, that lower metallicity environments typically host a population of smaller PAH grains. Currently, the only convincing model we have that explains this is the ``inhibited growth'' model\refedit{, where the average size of the modeled PAH population decreases with increasing metallicity and the overall mass of the PAHs declines} \citep{Whitcomb2024}.

\subsection{Radiation Field Effects} \label{sec:radiation}
While metallicity has a large impact on the PAH band ratio variations, the radiation field hardness still plays a key role. Metallicity gradients can contribute to the global trend of increasing PAH band ratios with galactocentric radius for many galaxies, but the four galaxies with star formation deserts, NGC\,1300, NGC\,1433, NGC\,1512, and NGC\,3351, show a\refedit{n} enhancement of the 3.3~$\micron$ relative to the 7.7 and 11.3~$\micron$ feature strengths in Figure \ref{fig:ratio_radius}. The location of these areas is shown in Figure \ref{fig:ratio_radius} between black dashed lines in these galaxies. In \citet{Baron2025}, the same areas were identified as having unusually high 11.3/7.7~$\micron$ PAH ratios, older stellar populations, and high starlight-to-dust ratios. The presence of the older stellar population suggests a softer radiation field heating the PAHs, resulting in a weaker 7.7~$\micron$ feature compared to 11.3~$\micron$ \citep{Draine2021, Chastenet2023, Baron2024}. 
  
\citet{Baron2024} used optical line ratios to break the degeneracy between the dependence of the 3.3/11.3~$\micron$ and 7.7/11.3~$\micron$ ratios on the size distribution and radiation field hardness. They calculated the relationship between the 3.3/11.3~$\micron$ PAH ratio and the [OIII]/H$\beta$, [NII]/H$\alpha$, and [SII]/H$\alpha$ ratios and found that the variation in radiation field must be accounted for to use this PAH ratio as a tracer of size distribution. To expand on their analysis of three galaxies to our sample of 19, Figure \ref{fig:optical_lines} shows the various optical line ratios compared to the 3.3/11.3 ratio in the entire sample of galaxies. The [OIII]/H$\beta$ and [SII]/H$\alpha$ and 3.3/11.3~$\micron$ ratio correlations are weakened by the presence of lower metallicity regions. Again, we see the effects of the enhancement of small PAHs traced by the 3.3~$\micron$ feature skewing the otherwise tight relation between the radiation field and band ratio. The trend with [NII]/H$\alpha$ remains because the low metallicity regions are all located at low values of [NII]/H$\alpha$, as expected from the correlation between [NII]/H$\alpha$ with metallicity at fixed ionization parameter. Figure \ref{fig:optical_lines_individual} shows that the differences in the dependence on different optical line ratios with PAH band ratios from \citet{Baron2024} are not from a difference in F335M continuum removal method, but from a wider range in radiation field properties from the inclusion of all 19 galaxies. 

When we color the BPT diagram by 3.3/11.3 and 7.7/11.3~$\micron$ ratios and metallicity in Figure \ref{fig:bpt}, we see two distinct trends between metallicity and radiation field hardness with PAH band ratio values. When metallicity is held at a constant high value, the radiation field has clear effects on both the 3.3/11.3 and 7.7/11.3~$\micron$ ratio values. The left panels of Figure \ref{fig:band_ratios} show the relatively high metallicity values in the LINER \refedit{and AGN} regions of the BPT diagram, while [NII]/H$\alpha$ spans $\sim$0.5 dex. When these diamond bins are created with and without AGN in Figure \ref{fig:bpt_grids}, we see that the main driver of the additional $\sim$0.2 dex observed variation in the 7.7/11.3~$\micron$ value is the shift toward neutral PAHs in the presence of AGN \refedit{possibly from the preferential destruction of ionized PAHs}
agreeing with several previous studies including \refedit{\citet{Garcia-Bernete2022}} and \citet{GarciaBernete2024}. \refedit{The variation in 7.7/11.3~$\micron$ ratio could also be caused by changing interstellar radiation field effects driven by the bulge and stellar components of these galaxies, not by the neutralization of PAHs by the AGN, as observed in \citet{Donnelly2024}.} 
When AGN hosts are removed from these bins, we see the radiation field effect both the 3.3/11.3 and 7.7/11.3~$\micron$ ratios, with enhanced shorter wavelength PAH features at harder radiation fields. The variation between star forming regions and LINER regions is $\sim$0.15 dex in 7.7/11.3~$\micron$ which is consistent with radiation field effects and agrees with the findings from \citet{Baron2025}. 

Figure \ref{fig:models_all} shows agreement with the \citet{Draine2021} model \refedit{grid coverage} and a clear dependence of 3.3/7.7 and 3.3/11.3~$\micron$ band ratio values on radiation field, traced by sSFR. \refedit{\citet{Draine2021} showed when} there is a change \refedit{towards harder radiation fields}, there is an enhancement of the 3.3~$\micron$ feature that is not caused by an intrinsic change in the PAH population properties, but by the enhancement of the feature strength from larger PAHs being heated by higher energy photons and emitting at shorter wavelengths. \refedit{This would cause a change in the PAH band ratios used as a tracer of size, such as 3.3/11.3~$\micron$, that would be similar to a shift towards a smaller PAH population. Changing the radiation field intensity parameter, $U$, in the models while holding charge, size, and radiation field hardness constant, yields no change in the PAH band ratios measured here \citep{Pathak2026}.}

\refedit{When using sSFR as a tracer of radiation field hardness, our observed PAH band ratios do not follow the direction of 3.3~$\micron$ enhancement expected by the model grids and mostly fall inside the grid illuminated by the oldest stellar population. \citet{dale2025} also measured these PAH band ratios in the same sample of 19 galaxies, and when compared to \citet{Draine2021} model grids from stellar populations with ages ranging from 3~Myr to 1000~Myr, the PAH band ratio values were most consistent with being heated by an older stellar population. This may be partially explained by uncertainty in the PAH heating models, especially in the 3.3~$\micron$ feature, as it was not covered by {\em Spitzer}, so its cross-section is less constrained than other features.  One of the main issues discussed in \citet{Draine2021} is that their models assume complete conversion of photon energy to vibrational emission, and for the higher energy photons primarily driving the 3.3~$\micron$ emission, it is not clear that this is always a good approximation. The models also do not contain contributions from the 3.4~$\micron$ aliphatic feature or PAH continuum (this is a $\sim10\%$ effect given the observed 3.4 $\micron$ feature strength). Another potential explanation, given the strong correlation of PAH emission and CO, is that we expect that some PAH emission may arise from locations where the illuminating radiation field has been attenuated by dust to some degree which could be enough to soften the radiation field to where we see in Figure \ref{fig:models_all}. In general, we do see that the lowest sSFR points often tend to fall in regions of the grid which may be explained by a harder radiation field spectrum, as might be expected \refedit{based on \citet{Baron2024} and \citet{Baron2025}}. There is a group of points with very low sSFR that stray from the coverage of the grid in both band ratios, with a tendency toward higher 3.3~$\micron$ relative to 7.7~$\micron$ feature strength that come from the same galaxies that have star formation deserts. \refedit{To visualize which galaxies contribute the low sSFR points, w}e show the individual galaxy panels in Appendix \ref{sec:models_individual}.}

\section{Summary and Conclusions} \label{sec:conclusion}
In this work, we refine and expand an empirical method to isolate the PAH-correlated emission in the JWST NIRCam F335M filter by allowing the slope ($B_{\rm PAH}$) of the F335M/F300M versus F360M/F300M color, which is needed to remove PAH contamination in F360M, to vary with environment. We find that this slope correlates with galaxy properties, particularly sSFR and [NII]/H$\alpha$, indicating that changes in the PAH-correlated emission in F360M are tied to star formation and the radiation field spectrum. We observe evidence \refedit{for} a change in PAH properties with metallicity, specifically on average smaller PAHs, traced by increased relative emission in the 3.3~$\micron$ feature \refedit{compared to the 7.7 and 11.3~$\micron$ features}, at low metallicities. We also see PAH band ratio trends with radiation field when metallicity is relatively high and constant, reflecting the influence of radiation field hardness on PAH feature strength. 

The main results of this paper are summarized:
\begin{enumerate}
    \item We present a new method to \refedit{to isolate the PAH emission at $\sim$3.3~$\micron$, including part of the 3.4~$\micron$ feature, 3.47~$\micron$ plateau, and PAH-correlated continuum by subtracting} the continuum from the F335M filter using the flanking F300M and F360M filters\refedit{. We apply this method to} a sample of 19 nearby galaxies on the star forming main sequence. In Section \ref{sec:method} we \refedit{provide} new  equations for continuum removal \refedit{based on} different observables, including sSFR, [NII]/H$\alpha$, and WISE4/WISE1. We find \refedit{a median} percent difference \refedit{of 6\%} when compared with the method presented by \citet{Sandstrom2023} due to the variation in $B_{\rm PAH}$. When compared to the method presented by \citet{Lai2020} that isolates only the 3.3~$\micron$ emission \refedit{and not all PAH-correlated emission in F335M}, we find \refedit{a median} percent difference \refedit{of 46\%} when accounting for the PAH contamination in the F360M filter. Spectral analysis is necessary to confirm what is being removed with F360M. 
    \item A change in the slope of PAH dominated emission in F335M/F300M vs F360M/F300M color space, $B_{\rm PAH}$, shows a dependence of PAH-correlated emission in the F335M and F360M bands on various galaxy properties explored in Section \ref{sec:bpah_meas}. The strongest correlations are with sSFR and optical line ratio [NII]/H$\alpha$. This indicates a dependence on the relative strength of PAH-correlated emission in the F335M and F360M filters on radiation field hardness and supports the idea that 3.4~$\micron$ carriers are destroyed in regions with higher intensities of star formation\refedit{, assuming the 3.4~$\micron$ feature is responsible for variations in the F360M PAH-correlated emission. Alternatively, other PAH-correlated contributions in F360M could be responsible, including the recently identified plateau feature \citep{Boersma2023}}.
    \item In general, the 3.3/11.3 and 3.3/7.7~$\micron$ band ratios increase with increasing galactocentric radius, with several galaxies showing higher band ratios compared to the sample-wide average shown in Figures \ref{fig:ratio_radius} and \ref{fig:ratio_radius2}. The enhanced PAH band ratio values in NGC\,1300, NGC\,1433, NGC\,1512, and NGC\,3351 discussed in Section \ref{sec:environmental_dependence} are in ``star formation deserts'' and are the same regions as those identified in \citet{Baron2025} and \citet{Pathak2026} as having ``anomalous PAH ratios'', where the 7.7~$\micron$ feature is suppressed relative to the 3.3 and 11.3~$\micron$ features. 
    \item Galaxies with \refedit{low metallicities} show \refedit{3.3/7.7 and 3.3/11.3~$\micron$} band ratio radial profiles above average, including IC\,5332, NGC\,2835, and NGC\,5068. When binned by metallicity in Figure \ref{fig:metallicity}, we see a significant correlation between decreasing values of both 3.3/7.7 and 3.3/11.3~$\micron$ band ratios with increasing metallicity across all galaxies. When we compare to models of PAHs and stellar populations at various metallicities covered in our galaxy sample in Section \ref{sec:size}, we see that the observed metallicity trend cannot be explained by the enhanced hardness of radiation fields for low-metallicity stellar populations alone. \refedit{We see this trend across a broad sample of 19 galaxies and rule out radiation field spectral changes as the cause. We attribute the band ratio trends to lower average PAH size in low metallicity conditions, agreeing with the ``inhibited growth'' model \citep{Whitcomb2024}.}
    \item We find a correlation of 3.3/11.3~$\micron$ ratio with optical line ratio [NII]/H$\alpha$ in Figure \ref{fig:optical_lines}. Restricting our analysis to the same galaxies as \citet{Baron2024} in Figure \ref{fig:optical_lines_individual}, shows good agreement \refedit{with that paper}, but when we expand to the full sample, the correlation weakens with [OIII]/H$\beta$ and [SII]/H$\alpha$ in galaxies \refedit{that reach low values of metallicity}. We observe a dependence on BPT diagram location with PAH band ratio in Figure \ref{fig:bpt}. \refedit{We find two distinct tracks of PAH band ratio evolution in the star-forming to LINER and AGN regimes of the BPT diagram, separating the effects of metallicity and radiation field hardness on these measurements.}
     \item We find that at fixed metallicity, radiation field hardness affects the 3.3/11.3 and 7.7/11.3~$\micron$ band ratios in Figure \refedit{\ref{fig:bpt}.} While the intrinsic properties of the PAH population are not changing, the radiation field heating the PAHs can increase the relative amount of 3.3 and 7.7~$\micron$ emission from the higher energy photons discussed in Section \ref{sec:radiation}. \refedit{This trend can be seen in the BPT diagram, but is sub-dominant to metallicity related trends when plotting the data in PAH band ratio space as in Figure~\ref{fig:band_ratios}.}
     \item We \refedit{compare to} the \citet{Draine2021} model grids in Figure \ref{fig:models_all}, showing the impact of stellar radiation fields on size and charge distributions of PAH populations in our galaxies. Binning by metallicity in regions identified as star forming, using the \citet{Kauffmann2003} line on the BPT diagram, 
    shows a PAH size dependence on metallicity with decreasing 3.3/11.3~$\micron$ ratio with increasing metallicity in Figure \ref{fig:bpt_grids}. \refedit{The decreasing values of 7.7/11.3~$\micron$ ratios observed in AGN and LINER regions with increasing [NII]/H$\alpha$ are possibly caused by a shift towards more neutral PAHs around AGN or a changing interstellar radiation field.}
    When we exclude AGN from these binned points, the observed shifts in both the 3.3/11.3 and 7.7/11.3~$\micron$ ratios can be explained changes by radiation field hardness.
   
\end{enumerate}
\indent In conclusion, we have developed a refined, robust prescription for isolating emission from the 3.3~$\micron$\refedit{, and part of the 3.4~$\micron$} feature\refedit{s, 3.47~$\micron$ plateau, and PAH-correlated continuum} in nearby galaxies. Our analysis of PAH diagnostics versus physical conditions and environments within galaxies is just scratching the surface of what can be done with \refedit{JWST F300M, F335M, and F360M imaging}. We expect these results to be useful in a wide variety of extragalactic contexts as the archive of JWST observations continues to grow.

\section*{Acknowledgments}
The authors thank the anonymous referee for their comments and suggestions that improved this paper. This work was carried out as part of the PHANGS collaboration. This work is based on observations made with the NASA/ESA/CSA James Webb Space Telescope. The data were obtained from the Mikulski Archive for Space Telescopes at the Space Telescope Science Institute, which is operated by the Association of Universities for Research in Astronomy, Inc., under NASA contract NAS 5-03127 for JWST. These observations are associated with programs 2107 and 3707. Support for programs 2107 and 3707 were provided by NASA through a grant from the Space Telescope Science Institute, which is operated by the Association of Universities for Research in Astronomy, Inc., under NASA contract NAS 5-03127. The specific observations can be accessed via doi:\href{https://archive.stsci.edu/doi/resolve/resolve.html?doi=10.17909/ew88-jt15}{10.17909/ew88-jt15}.

H.K., K.S., L.H., and M.P. acknowledge funding support from JWST-GO-02107.006-A and JWST-GO-03707.005-A. H.K. and K.S. acknowledge funding support from JWST-GO-04793.005-A. H.K. thanks Ilyse Clark, Beck Dacus, Ryan Vaught, \refedit{and Fergus Donnan} for insightful conversations that greatly improved this work.

JC acknowledges funding from the Belgian Science Policy Office (BELSPO) through the PRODEX project “JWST/MIRI Science exploitation” (C4000142239). OE gratefully acknowledges funding from the Deutsche Forschungsgemeinschaft (DFG, German Research Foundation), project 541068876. DP acknowledges support from the NSF GRFP. ER acknowledges the support of the Natural Sciences and Engineering Research Council of Canada (NSERC), funding reference number RGPIN-2022-03499 and the support of the Canadian Space Agency, funding reference 23JWGO2A07. MB acknowledges support by the ANID BASAL project FB210003 and by the French government through the France 2030 investment plan managed by the National Research Agency (ANR), as part of the Initiative of Excellence of Université Côte d’Azur under reference number ANR-15-IDEX-01. This research was funded, in whole or in part, by the French National Research Agency (ANR), grant ANR-24-CE92-0044 (project STARCLUSTERS). RSK and SCOG acknowledge financial support from the European Research Council via the ERC Synergy Grant ``ECOGAL'' (project ID 855130),  from the German Excellence Strategy via the Heidelberg Cluster of Excellence (EXC 2181 - 390900948) ``STRUCTURES'', and from the German Ministry for Economic Affairs and Climate Action in project ``MAINN'' (funding ID 50OO2206). RSK thanks for computing resources provided by the Ministry of Science, Research and the Arts (MWK) of the State of Baden-W\"{u}rttemberg through bwHPC and the German Science Foundation (DFG) through grants INST 35/1134-1 FUGG and 35/1597-1 FUGG, and also for data storage at SDS@hd funded through grants INST 35/1314-1 FUGG and INST 35/1503-1 FUGG. HAP acknowledges support from the National Science and Technology Council of Taiwan under grant 110-2112-M-032-020-MY3 and 113-2112-M-032 -014 -MY3. A.K.L. and D.P. gratefully acknowledge support from NSF AST AWD 2205628, JWST-GO-02107.009-A, and JWST-GO-03707.001-A. A.K.L. also gratefully acknowledges support by a Humbolt Research Award. 

This research made use of  NASA's Astrophysics Data System, \texttt{NumPy} \citep{harris2020array}, \texttt{SciPy} \citep{Virtanen_2020}, \texttt{Astropy}, a community-developed core Python package for Astronomy \citep{2018AJ....156..123A, 2013A&A...558A..33A}, \texttt{matplotlib}, a Python library for publication quality graphics \citep{Hunter:2007}, and Scientific Colour Maps \citep{crameri2023}. 

\appendix

\section{Image Processing} \label{sec:more_processing}
We briefly describe several important aspects of data reduction for creating the F335M$_{\rm PAH}$ maps, including: corrections of astrometric offsets between the F300M, F335M, and F360M filters and anchoring the maps to obtain accurate zero-points; and convolution to matched PSFs.

Early attempts of continuum subtraction revealed \refedit{sub-pixel shifts in some images} between the F300M, F335M, and F360M filters remaining after processing data with the JWST pipeline. To fix this offset, we match the F335M and F360M filters to the astrometry of the F300M. We first calculate the median background and detect point sources above it with SNR $>5$ in both images using the \texttt{Photutils Background2D} and \texttt{SourceFinder} functions. After making a source catalog, we find the indices of matching point sources within a specified distance, 0\farcs2, and iteratively remove outliers with sigma-clipping. The distances between these matching point sources are measured, and the median value of the offsets is used as the final offset between the images. The typical sizes of these offsets range between $\sim$ 0.005--0.05 pixel shifts. More details on this analysis will be published by Weinbeck et al. in prep.

Before removing the continuum in F335M, we checked for and corrected any absolute flux calibration offsets in a similar method to that presented by \citet{Leroy2023}. The coverage of NIRCam and MIRI observations of the PHANGS galaxies does not include much background off the galaxy to measure a zero-level, making it difficult to correct for any background offsets. We found that several of these galaxies have small offsets with groups of either strictly positive or negative points in places we expect to be at the noise limit. To correct this, we use WISE1 observations to anchor the background because of their wider field of view. First, the JWST data were convolved to the same resolution and put on the same pixel grid as WISE1 data at 7\farcs5 resolution. We then fit a line between the JWST F335M filter and the external data, providing an intercept value that gives the offset. We only use pixels with values between the 5th and 40th percentiles, with an upper limit of 0.8 MJy sr$^{-1}$ on the NIRCam data to determine this fit. After completing this process for all galaxies in the F335M filter, we then anchor the F300M and F360M internally in a similar way at 1$''$ resolution, with 0\farcs5 pixels. This process, done with \texttt{PJPipe}, is explained in greater detail in \citet{williams2024} and \citet{Chown2025a}.

\begin{figure}[ht]
      \centering
      \includegraphics[width = 6 in]{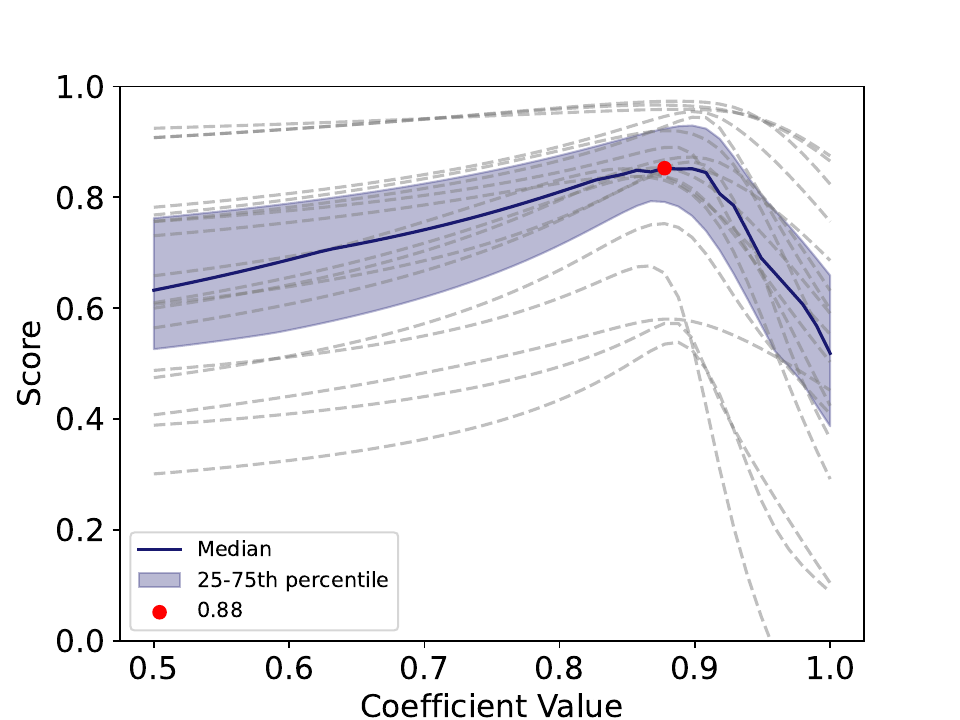}
      \caption{Each gray curve shows the score, a combination of the correlation coefficient between the original F335M$_{\rm PAH}$ map made with all three medium band filters and the F335M$_{\rm PAH}$ map made with F335M and F300M only with that coefficient value using Equation \ref{eq:cycle2} with a negative fraction penalty to reduce over-subtraction. The dark blue line shows the median score at each coefficient value while the shaded region shows the 25th--75th percentile. The coefficient with the highest median score, 0.88, is shown in red.}
      \label{fig:cycle2}
  \end{figure}

\section{Galaxies Without F360M} \label{sec:without_360}
The PHANGS galaxies observed in Cycle 2 \citep[GO 3707: PI Leroy,][]{Chown2025a} were reduced with the same pipeline presented by \citet{williams2024}, but have different filter coverage than those observed in Cycle 1. The galaxies observed in Cycle 2 include F150W, F187N, F300M, F335M, F770W, and F2100W observations. The lack of observations in the F360M filter changes how we perform the F335M continuum subtraction. To use the nearby available medium band filter, F300M, we calculate subtraction coefficients for each based on the results from the Cycle 1 data. To calculate these coefficients, we create F335M$_{\rm PAH}$ maps from Cycle 1 galaxies using the Equation 
\begin{equation} \label{eq:cycle2}
    \mathrm{F335M}_{\rm PAH} = \mathrm{F335M} - (C \times \mathrm{F300M})
\end{equation}
for a range of $C$ values. We mask the typical stripe value in the maps, 0.04 ~MJy~sr$^{-1}$, to only measure the correlation above the noise. The best coefficient value was determined by maximizing the correlation between the resulting PAH map without F360M and the one created using the method presented in this work. Coefficients were calculated for all Cycle 1 galaxies, and the median was used as the final value. This gives the resulting $\mathrm{F335M}_{\rm PAH}$ Equation for the Cycle 2 PHANGS galaxies
\begin{equation}
    \mathrm{F335M}_{\rm PAH} = \mathrm{F335M} - (0.88\cdot \mathrm{F300M}). 
\end{equation}
 To test the accuracy of these maps in a real application, we measure the 3.3/7.7~$\micron$ ratio using the maps made with F360M and without. We find an average percent difference of the band ratio across all the galaxies of 12\%, with average galaxy-wide deviations up to 38\%.  
  \section{F335M$_{\text{PAH}}$ and Continuum Maps} \label{sec:maps}
  Here we show the F335M$_{\text{PAH}}$ and continuum maps for each of the 19 galaxies. Striping can be seen across several of the maps, with typical values of $\pm$0.04 ~MJy~sr$^{-1}$. 
 \begin{figure*}[h!]
      \centering
      \includegraphics[width = \textwidth]{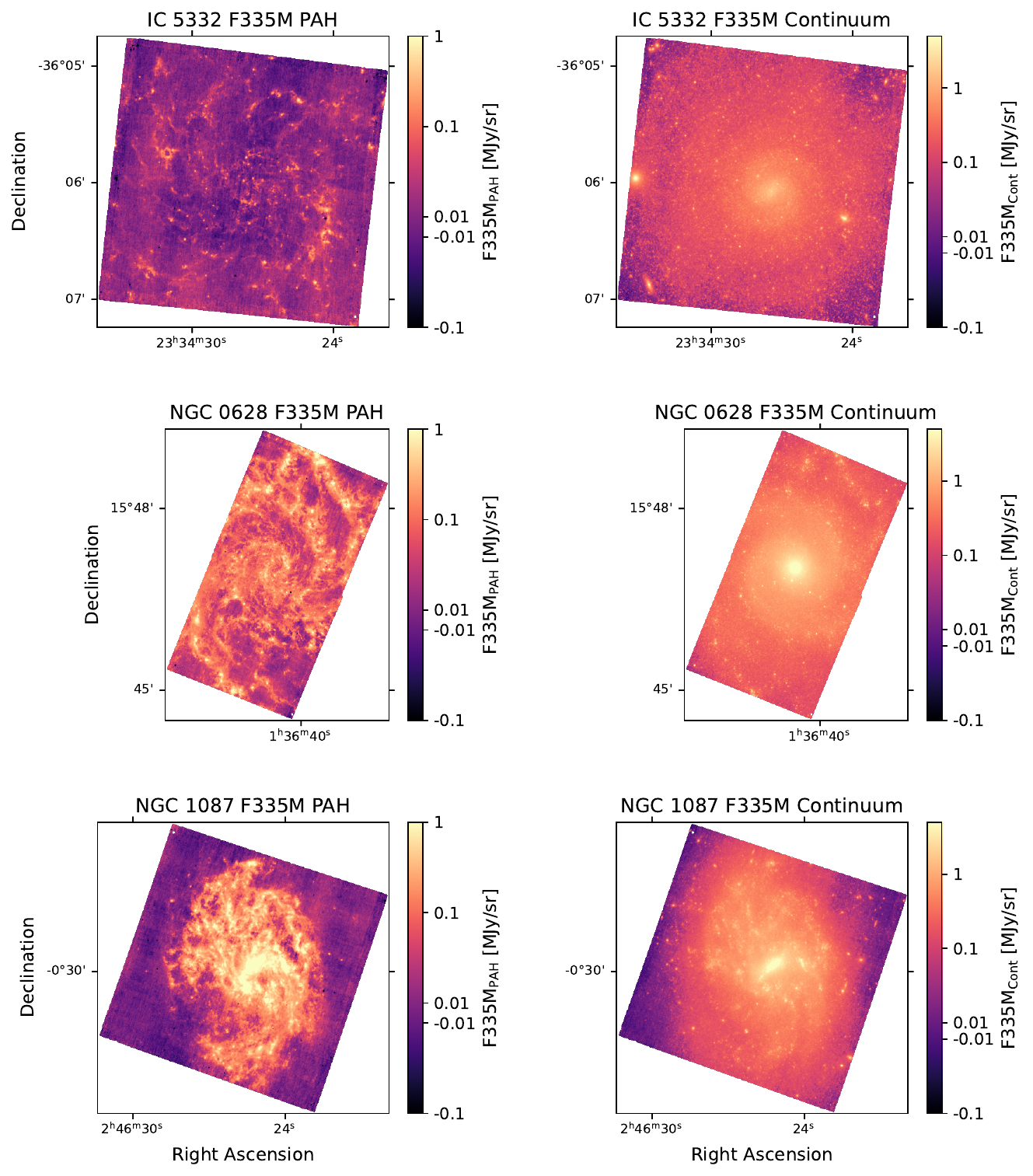}
      \caption{F335M$_{\rm PAH}$ maps for each galaxy in the PHANGS JWST Cycle 1 Survey using the individually calculated galaxy-wide values shown as open faced points in Figure \ref{fig:megatable_regions} and presented in Table \ref{tab:table}.}
      \label{fig:all_galaxies}
  \end{figure*}
   \begin{figure*}[h!]
      \centering
      \includegraphics[width = \textwidth]{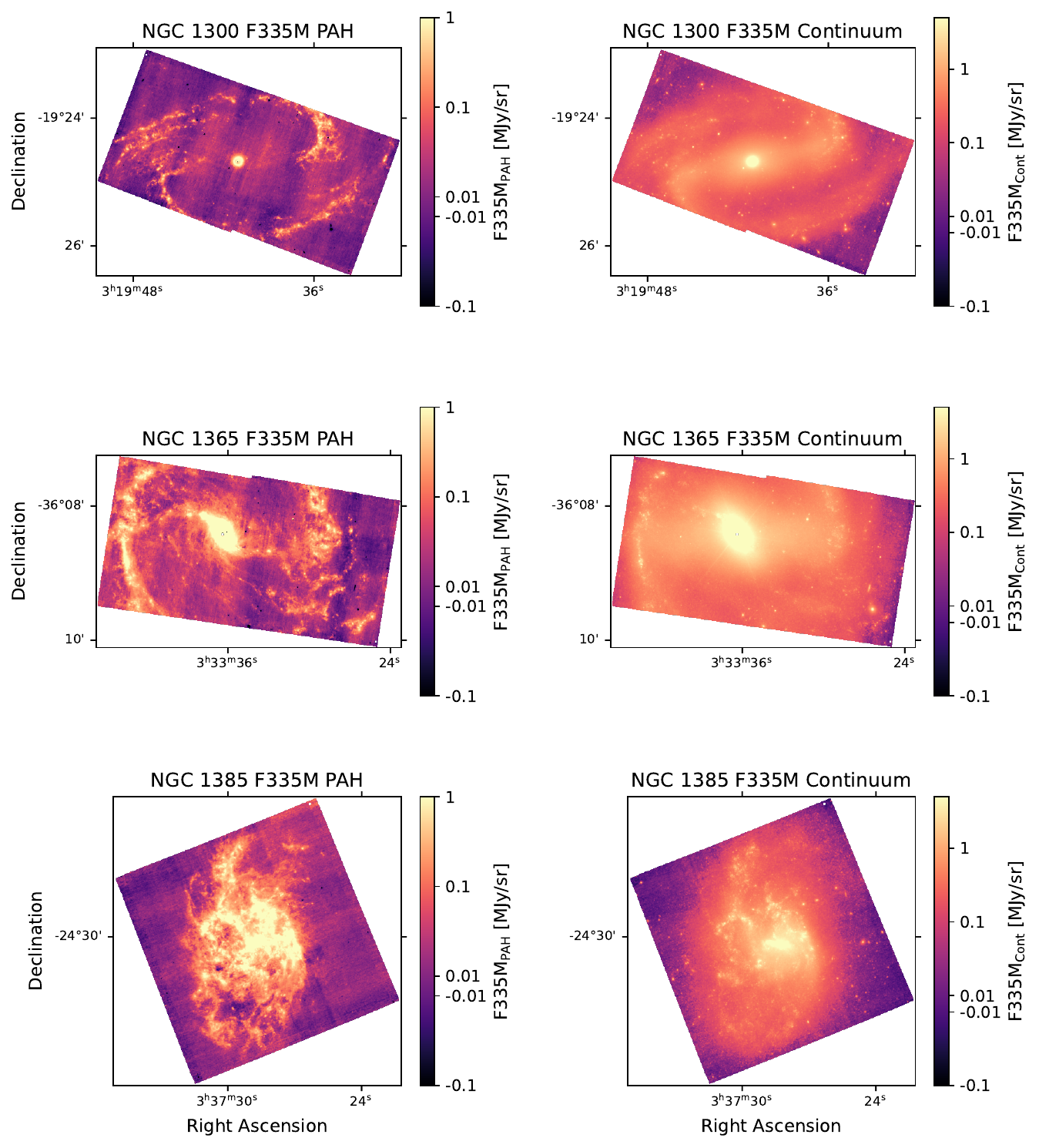}
      \caption{F335M$_{\rm PAH}$ maps for more galaxies using the individually calculated galaxy-wide values shown as open faced points in Figure \ref{fig:megatable_regions} and presented in Table \ref{tab:table}.}
      \label{fig:all_galaxiesb}
  \end{figure*}
  \begin{figure*}[h!]
      \centering
      \includegraphics[width = \textwidth]{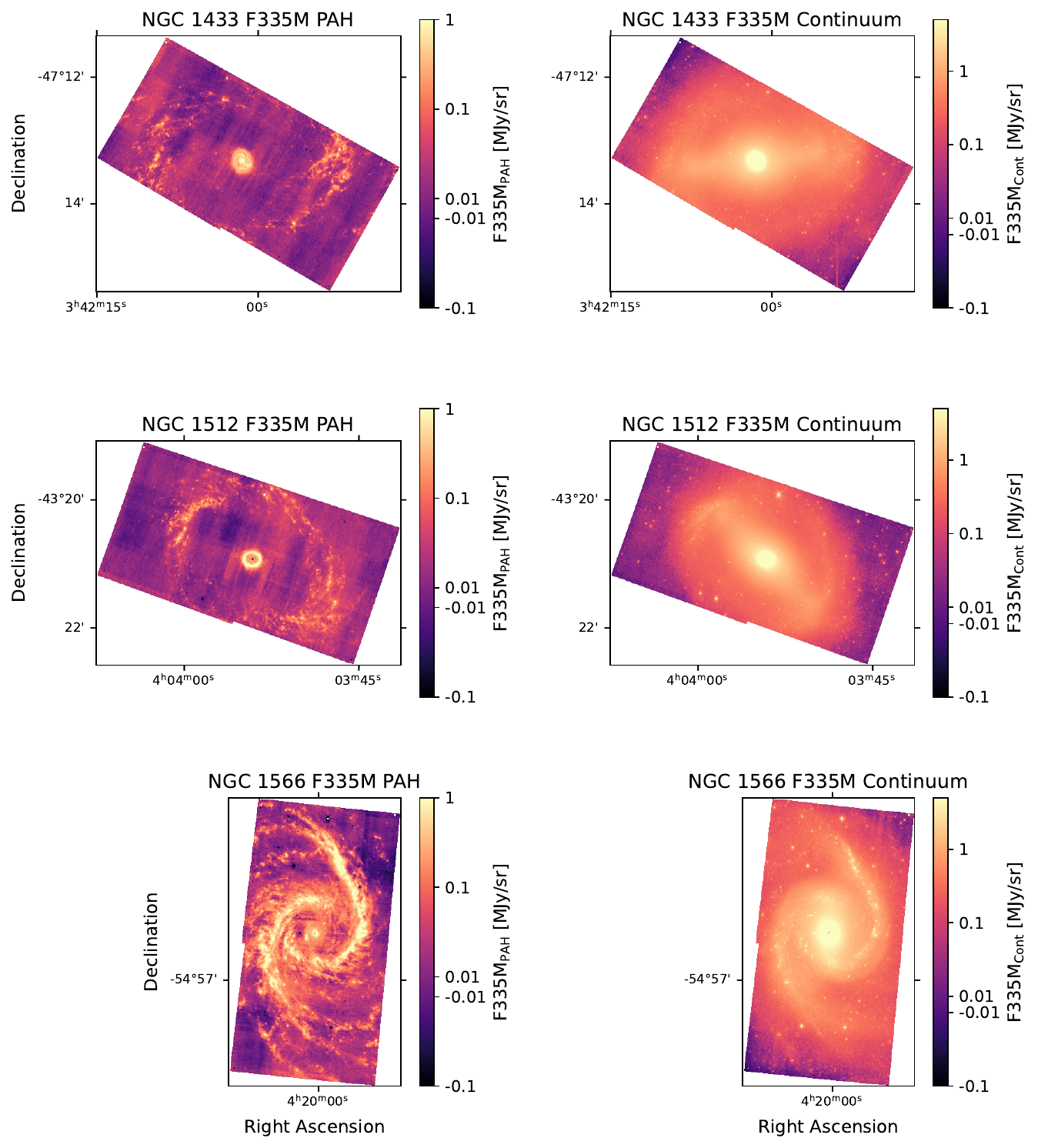}
      \caption{F335M$_{\rm PAH}$ maps for more galaxies using the individually calculated galaxy-wide values shown as open faced points in Figure \ref{fig:megatable_regions} and presented in Table \ref{tab:table}.}
      \label{fig:all_galaxiesc}
  \end{figure*}
  \begin{figure*}[h!]
      \centering
      \includegraphics[width = \textwidth]{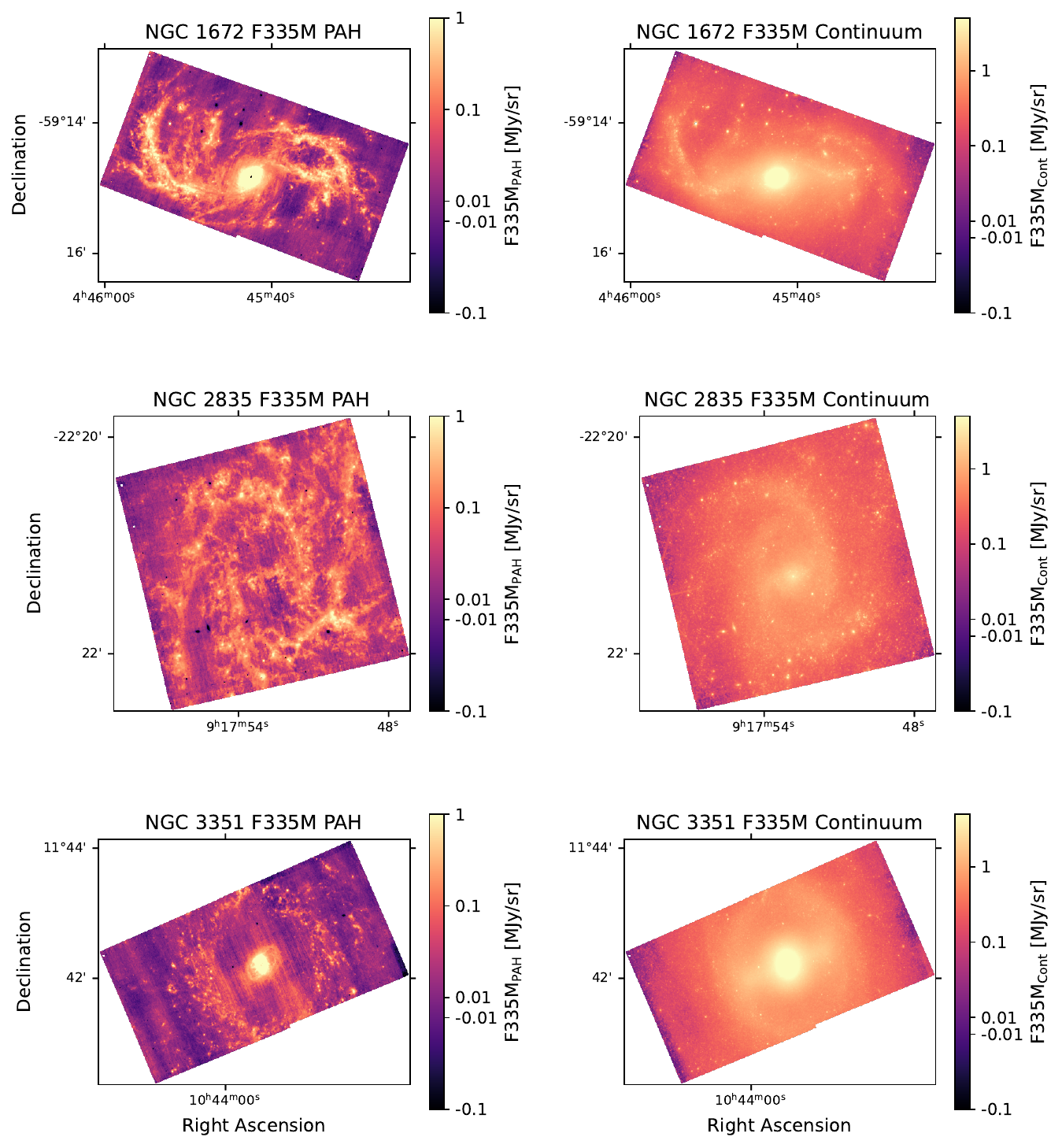}
      \caption{F335M$_{\rm PAH}$ maps for more galaxies using the individually calculated galaxy-wide values shown as open faced points in Figure \ref{fig:megatable_regions} and presented in Table \ref{tab:table}.}
      \label{fig:all_galaxiesd}
  \end{figure*}
  \begin{figure*}[h!]
      \centering
      \includegraphics[width = \textwidth]{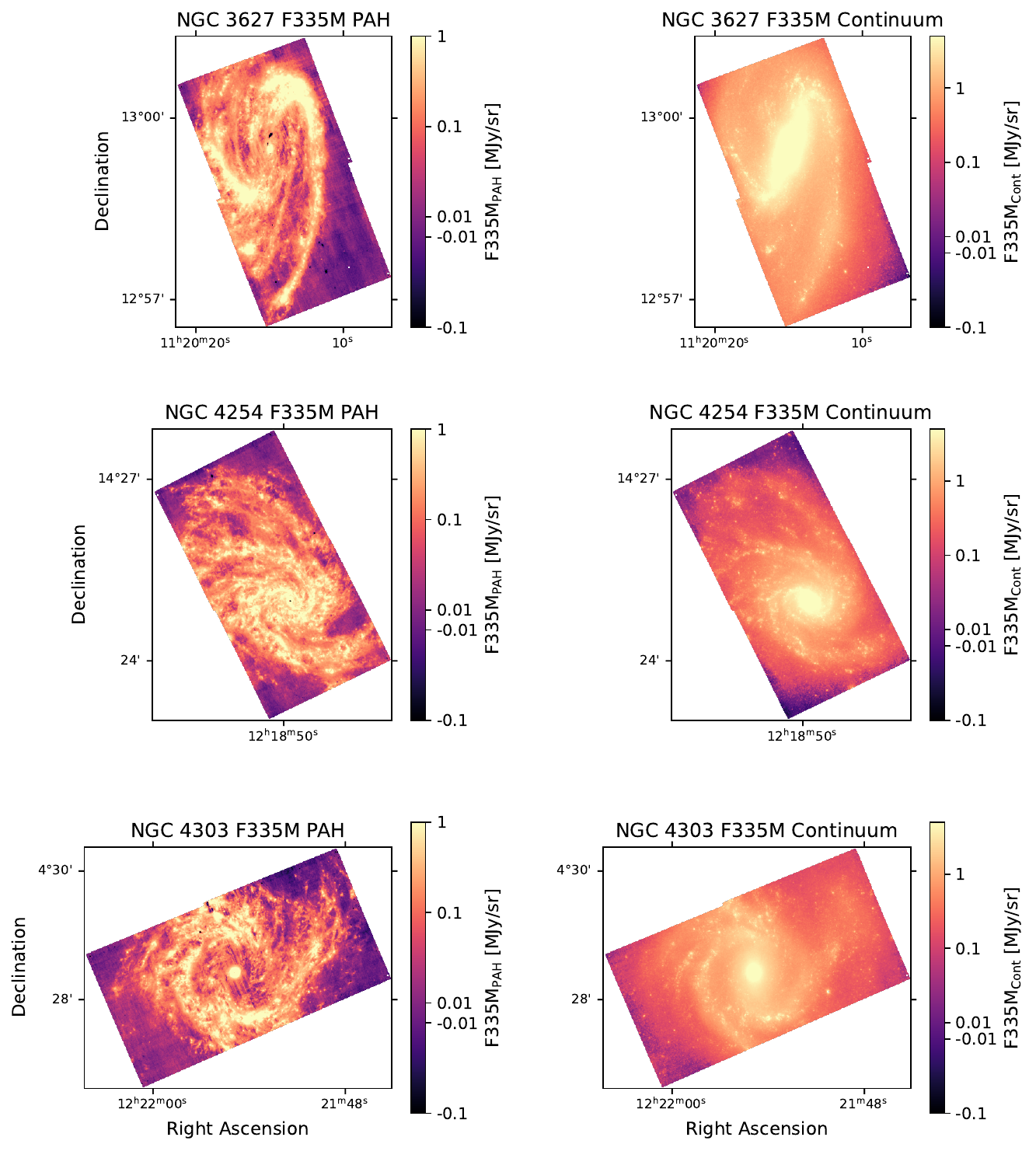}
      \caption{F335M$_{\rm PAH}$ maps for more galaxies using the individually calculated galaxy-wide values shown as open faced points in Figure \ref{fig:megatable_regions} and presented in Table \ref{tab:table}.}
      \label{fig:all_galaxiese}
  \end{figure*}
  \begin{figure*}[h!]
      \centering
      \includegraphics[width = \textwidth]{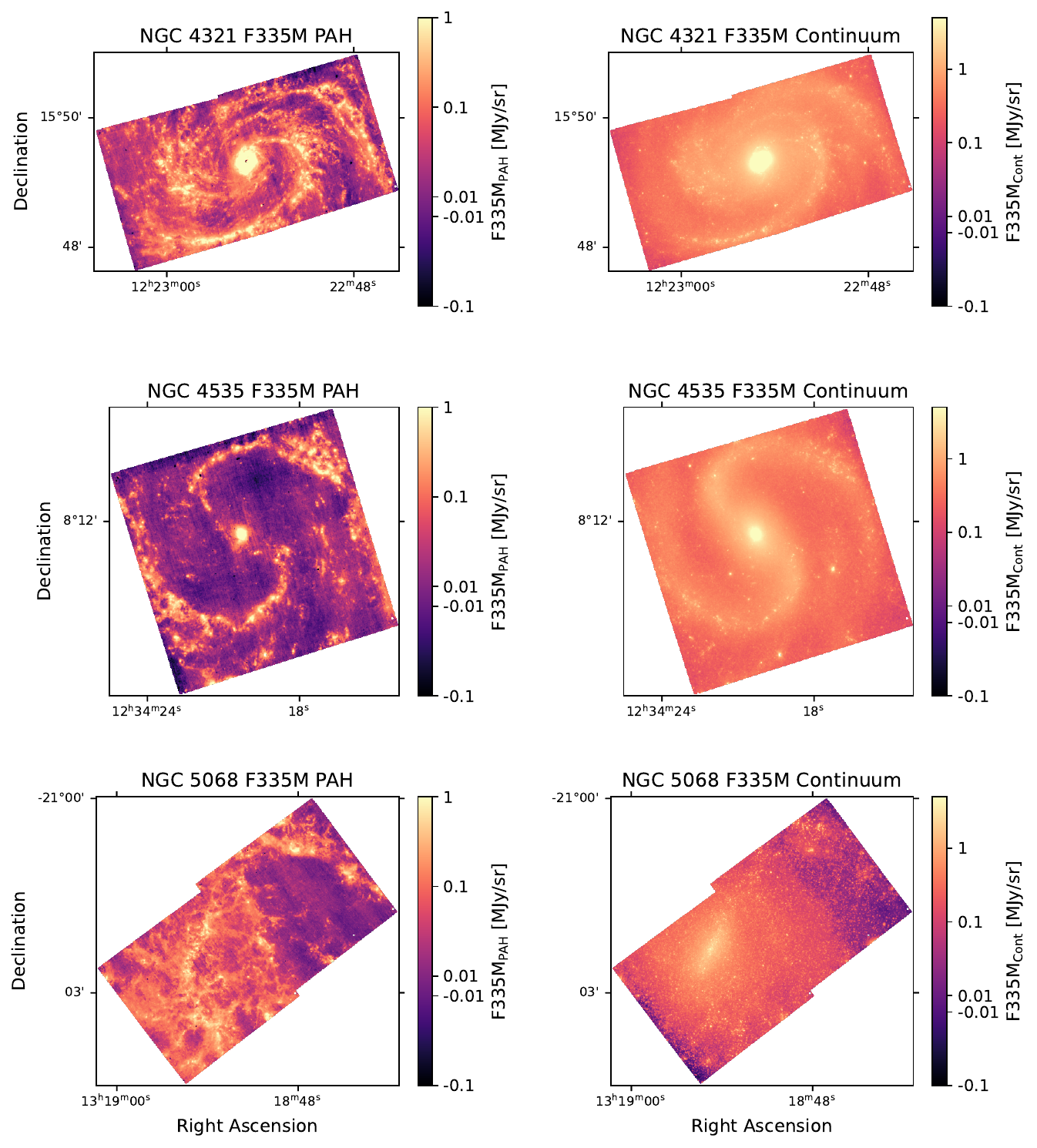}
      \caption{F335M$_{\rm PAH}$ maps for more galaxies using the individually calculated galaxy-wide values shown as open faced points in Figure \ref{fig:megatable_regions} and presented in Table \ref{tab:table}.}
      \label{fig:all_galaxiesf}
  \end{figure*}
  \begin{figure*}[h!]
      \centering
      \includegraphics[width = \textwidth]{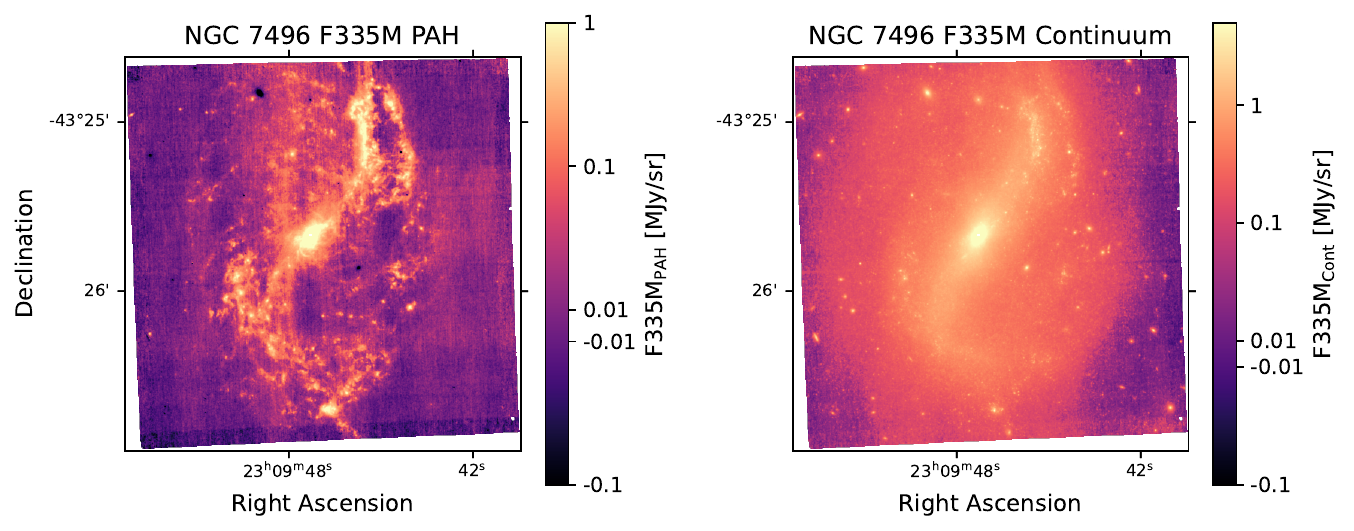}
      \caption{F335M$_{\rm PAH}$ maps for more galaxies using the individually calculated galaxy-wide values shown as open faced points in Figure \ref{fig:megatable_regions} and presented in Table \ref{tab:table}.}
      \label{fig:all_galaxiesg}
  \end{figure*}
\clearpage
  
  \section{Individual Galaxies on Draine et al. 2021 Grids}\label{sec:models_individual}
\begin{figure*}[h!]
      \centering
      \includegraphics[width = \textwidth]{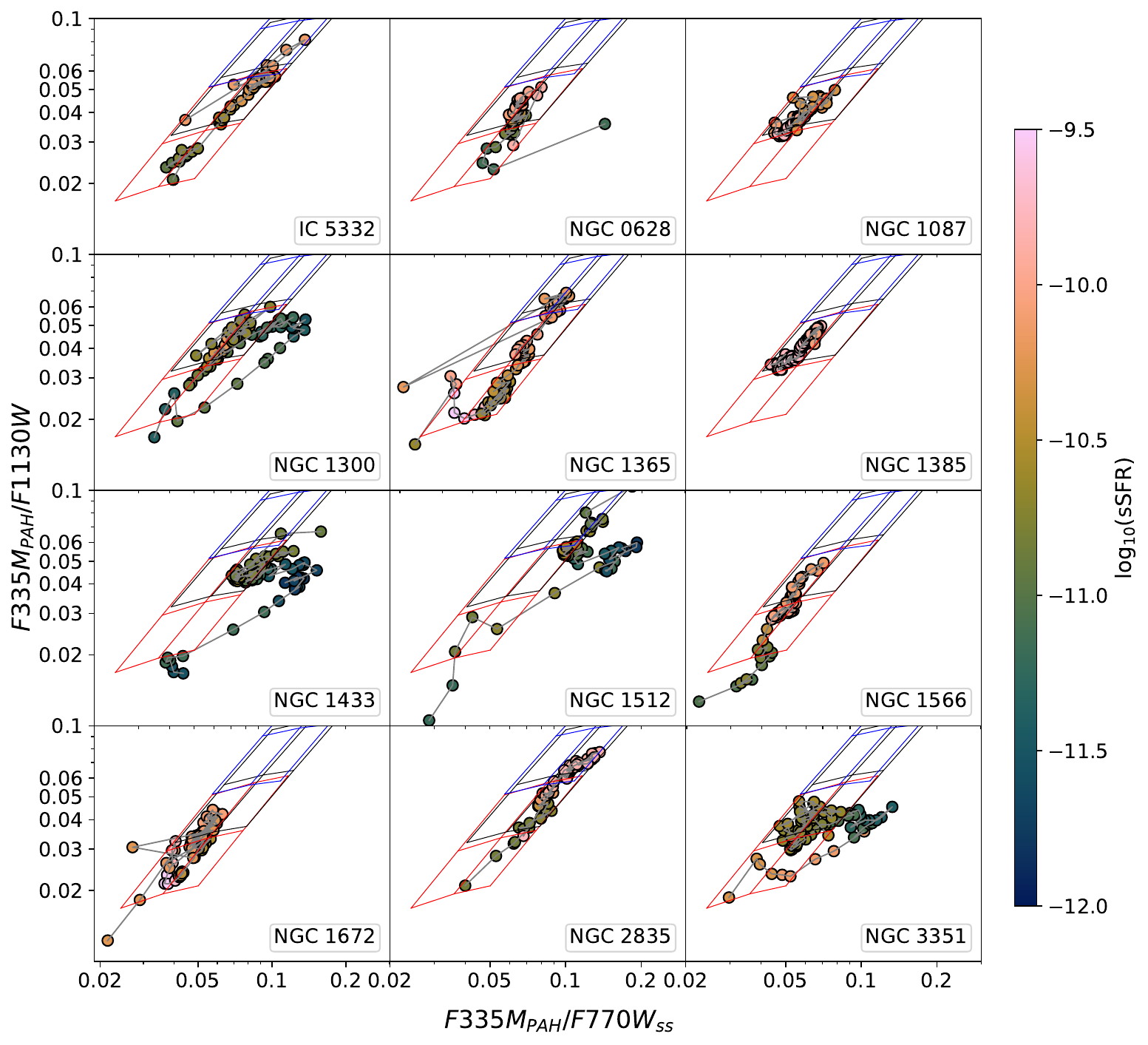}
      \caption{The same points as Figure \ref{fig:models_all}, binned in the same way as Figures \ref{fig:ratio_radius} and \ref{fig:ratio_radius2}, with bin sizes 0.01 $r_{25}$, separated by galaxy. The color bar shows sSFR, while the gray line connects each point going from the center of the galaxy outwards. Dust model grids from varying radiation fields, including the M31 bulge \citep{Groves2012}, mMMP \citep{mathis1983}, and BC03 3~Myr old starburst \citep{Bruzual2003} from \citet{Draine2021}, are shown in red, black, and blue, respectively. Small, standard, and large labels refer to three PAH size populations modeled in \citet{Draine2021}, while high, standard, and low label the three ionization fractions modeled.}
      \label{fig:models_individual}
  \end{figure*}

  Figure \ref{fig:models_individual} shows the same grids shown in Figure \ref{fig:models_all} with individual panels for each galaxy. The color of the points shows sSFR and a gray line connects the inner to outer points of each galaxy. When considering each galaxy individually, NGC\,1433, NGC\,1512, NGC\,3351, and NGC\,3627 are easily identified as the hosts for \refedit{the outliers in Figure \ref{fig:models_all}.} Several of these are the same galaxies identified in \citet{Baron2025} and \citet{Pathak2026} as having anomalous PAH ratios and show differences in band ratio trends with galactocentric radius in their star formation deserts in Figures \ref{fig:ratio_radius} and \ref{fig:ratio_radius2}. The lowest sSFR points \refedit{shift} towards the larger size and lower ionization models.

  \begin{figure*}[h!]
      \centering
      \includegraphics[width = \textwidth]{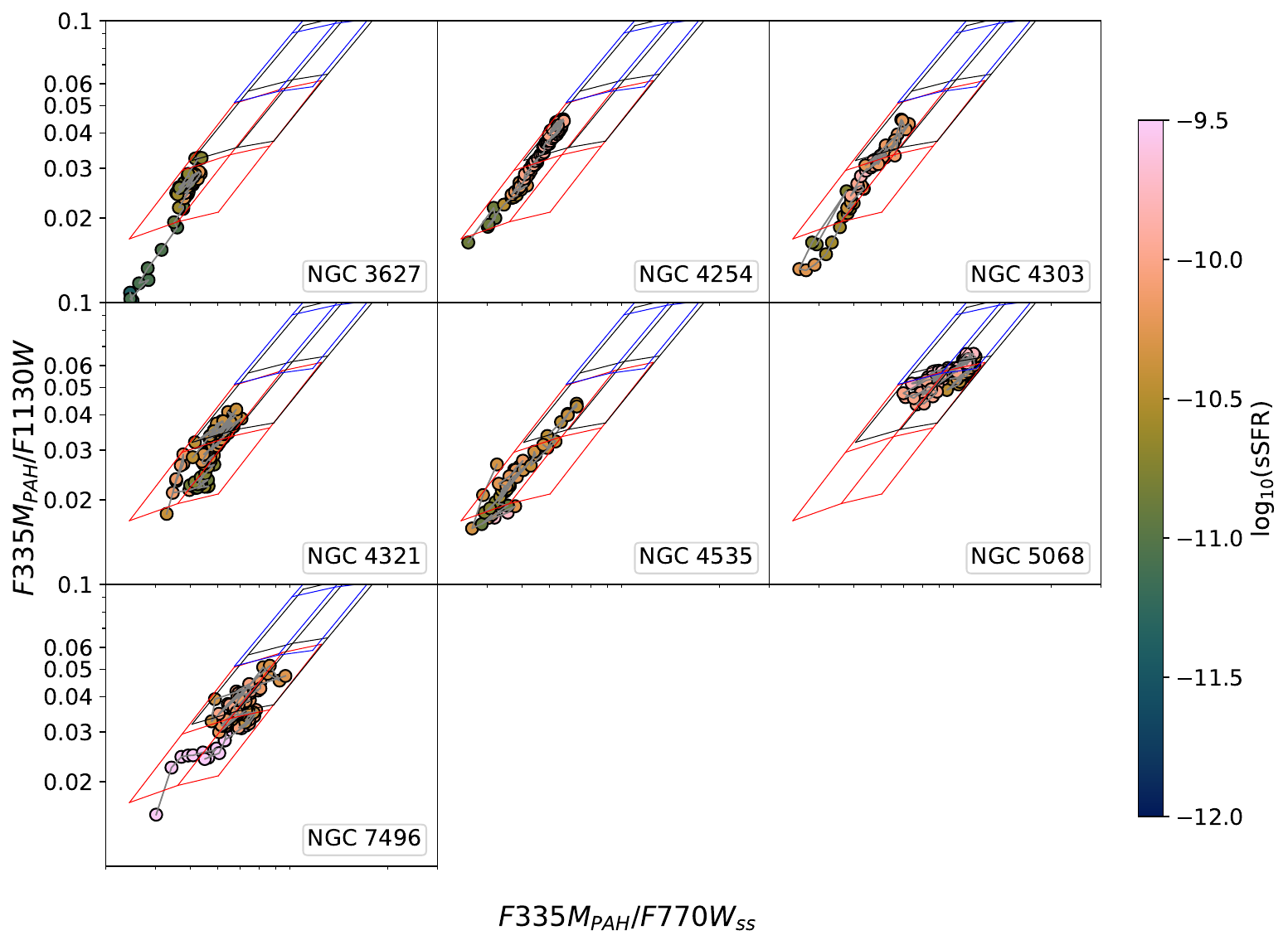}
      \caption{Figure \ref{fig:models_individual} (Continued).}
      \label{fig:models_individual2}
  \end{figure*} 

\section{\refedit{3.4~$\micron$ contribution to F360M}}
\label{sec:3p4}
\refedit{To assess the contribution of the 3.4~$\micron$ feature to F360M and our $B_{\rm{PAH}}$ measurement, we used an example spectrum from the Whirlpool Treasury in M51 (GO 3435; PI Sandstrom \& Dale). We first decomposed the spectrum using PAHFIT \citep{Smith2007} to measure the integrated intensities of the PAH features. We include Drude profiles for several PAH features in the relevant wavelength range following the decomposition method from PDRs4All \citep{Chown2024}. The 3.3~$\micron$ feature is modeled by two components with central wavelengths 3.2 and 3.29~$\micron$; 3.4~$\micron$ is modeled as three components with central wavelengths 3.39, 3.4, and 3.42; the 3.47~$\micron$ feature is one component; and some of the features comprising the PAH-correlated continuum are modeled as Drude profiles with central wavelengths 3.51 and 3.56. We used the results from PAHFIT for a single pixel in M51 to create a PAH and aliphatic-only spectrum with realistic FWHM and integrated intensity values. We use the throughput curves for F335M and F360M to test the effects of changing the sum of the three 3.4~$\micron$ feature strengths on the ratio of F335M/F360M. We calculate the values of synthetic photometry in two cases, first using the baseline results from PAHFIT, and second, scaling up the three features at 3.4~$\micron$ to double their total integrated intensity. The ratio of the synthetic photometry of F335M/F360M decreases by 0.2, showing that it contributes a measurable amount to the F360M filters and in our measurements of $B_{\rm{PAH}}$. This is larger than the errors calculated on $B_{\rm{PAH}}$ presented in Table \ref{tab:table}.}

\section{\refedit{Additional PAH Band Ratios Separated by Environment}}
\begin{figure*}[ht]
      \centering
      \includegraphics[width = \textwidth]{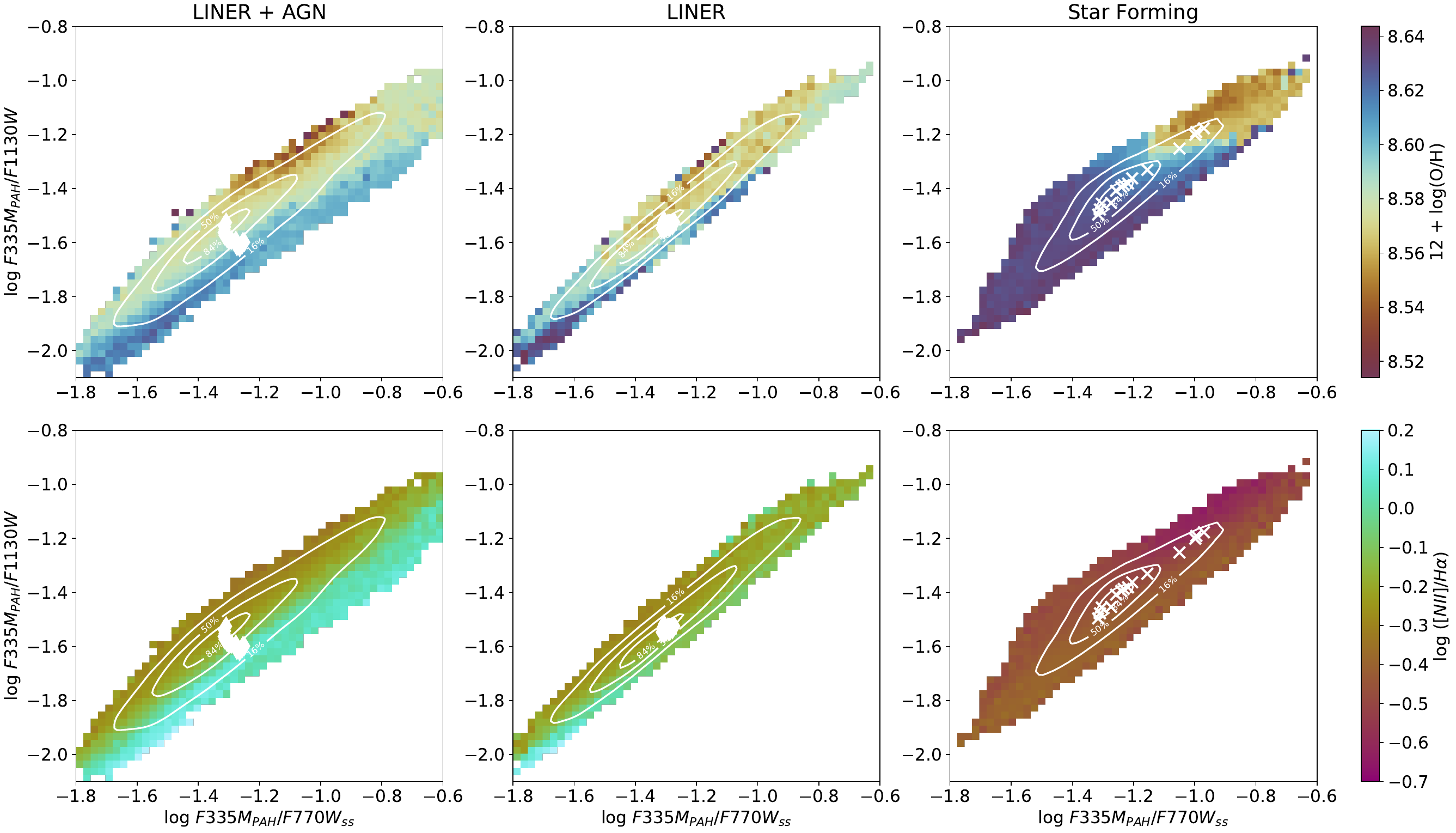}
      \caption{\refedit{3.3/11.3 plotted against 3.3/7.7 at 150~pc resolution colored by metallicity (top) and [NII]/H$\alpha$ (bottom) for comparison with Figure \ref{fig:models_all}. We isolate the points that are considered LINERs and AGN (left), LINERs (middle), and star-forming (right). Contours represent the 16th, 50th, and 84th percentiles. In the star forming regions we see the expected enhancement of the 3.3~$\micron$ feature relative to 7.7 and 11.3~$\micron$ with decreasing metallicity. In the LINER and AGN and LINER only regions, we see variations of the 3.3/11.3 and 3.3/7.7~$\micron$ ratios that are primarily correlated with metallicity.  In this projection of the data, [NII]/H$\alpha$ also tracks metallicity, resulting in similar gradients across band ratio space.}}
      \label{fig:band_ratios2}
  \end{figure*}

  \refedit{We recreate Figure \ref{fig:band_ratios} with the PAH band ratios 3.3/7.7 and 3.3/11.3~$\micron$ as axes in Figure \ref{fig:band_ratios2} to further investigate their correlation with metallicity and [NII]/H$\alpha$. We separate the environments into star-forming, LINER, and LINER and AGN regions. We clearly see the same correlation between band ratios and metallicity in the star-forming regions as previously discussed, enhanced 3.3~$\micron$ emission relative to the longer wavelength features at lower metallicity. The LINER and AGN and LINER only regions show the relationship between metallicity and [NII]/H$\alpha$, similarly to Figure \ref{fig:band_ratios}. To see the PAH band ratio dependence on the radiation field spectrum and metallicity, the data must be binned in BPT space, such as in Figure \ref{fig:bpt}. \citet{Draine2021} showed that there is a correlation between the 3.3/7.7 and 3.3/11.3~$\micron$ band ratios, and that the 3.3/7.7~$\micron$ band ratio is as sensitive to changes in the PAH and radiation field properties compared to 3.3/11.3 or 7.7/11.3~$\micron$.
  }

\clearpage
\bibliography{main}{}
\bibliographystyle{aasjournalv7}
\end{document}

%% file: affiliations.tex
\newcommand{\UCSD}{Department of Astronomy and Astrophysics, University of California, San Diego, CA 92093, USA}
\newcommand{\Carnegie}{The Observatories of the Carnegie Institution for Science, 813 Santa Barbara Street, Pasadena, CA 91101, USA}
\newcommand{\KIPAC}{Kavli Institute for Particle Astrophysics \& Cosmology (KIPAC), Stanford University, CA 94305, USA}
\newcommand{\OSU}{Department of Astronomy, The Ohio State University, Columbus, OH 43210, USA}
\newcommand{\Whitman}{Whitman College, 345 Boyer Avenue, Walla Walla, WA 99362, USA}
\newcommand{\UHeidelberg}{Astronomisches Rechen-Institut, Zentrum f\"ur Astronomie der Universit\"at Heidelberg, M\"onchhofstr. 12-14, D-69120 Heidelberg, Germany}
\newcommand{\UWyoming}{Department of Physics and Astronomy, University of Wyoming, Laramie, WY 82071, USA}
\newcommand{\Algoma}{Faculty of Computer Science \& Technology, Algoma University, Sault Ste. Marie, ON
P6A 2G4, Canada}
\newcommand{\UGent}{Sterrenkundig Observatorium, Universiteit Gent, Krijgslaan 281 S9, B-9000 Gent, Belgium}
\newcommand{\Columbus}{Center for Cosmology and Astroparticle Physics, 191 West Woodruff Avenue, Columbus, OH 43210, USA}
\newcommand{\uoa}{Department of Physics, University of Arkansas, 226 Physics Building, 825 West Dickson Street, Fayetteville, AR 72701, USA}
\newcommand{\UConn}{Department of Physics, University of Connecticut, 196A Auditorium Road, Storrs, CT 06269, USA}
\newcommand{\UAlberta}{Department of Physics, University of Alberta, Edmonton, AB T6G 2E1, Canada}
\newcommand{\ITA}{Universit\"{a}t Heidelberg, Zentrum f\"{u}r Astronomie, Institut f\"{u}r Theoretische Astrophysik, Albert-Ueberle-Str.\ 2, 69120 Heidelberg, Germany}
\newcommand{\IWR}{Universit\"{a}t Heidelberg, Interdisziplin\"{a}res Zentrum f\"{u}r Wissenschaftliches Rechnen, Im Neuenheimer Feld 225, 69120 Heidelberg, Germany}
\newcommand{\JHU}{\affiliation{Department of Physics and Astronomy, The Johns Hopkins University, Baltimore, MD 21218 USA}}
\newcommand{\Tamkang}{Department of Physics, Tamkang University, No.151, Yingzhuan Road, Tamsui District, New Taipei City 251301, Taiwan}
\newcommand{\Azur}{Université Côte d’Azur, Observatoire de la Côte d’Azur, CNRS, Laboratoire Lagrange, 06000, Nice, France}
\newcommand{\JBCA}{UK ALMA Regional Centre Node, Jodrell Bank Centre for Astrophysics, Department of Physics and Astronomy, The University of Manchester, Oxford Road, Manchester M13 9PL, UK}
\newcommand{\AIP}{Leibniz-Institut for Astrophysik Potsdam (AIP), An der Sternwarte 16, 14482 Potsdam, Germany}
\newcommand{\STScI}{Space Telescope Science Institute  (STScI),  Baltimore, MD, USA}
\newcommand{\ESOcl}{European Southern Observatory (ESO), Alonso de Córdova 3107, Casilla 19, Santiago 19001, Chile} 
\newcommand{\NOIRLab}{International Gemini Observatory/NSF NOIRLab, 950 N Cherry Ave, Tucson, AZ 85719, USA} 
\newcommand{\STScIESA}{\affiliation{AURA for the European Space Agency (ESA), Space Telescope Science Institute, 3700 San Martin Drive, Baltimore, MD 21218, USA}}
\newcommand{\umd}{Department of Astronomy University of Maryland College Park, MD 20742 USA}

\newcommand{\MPE}{Max-Planck-Institut f\"{u}r extraterrestrische Physik, Giessenbachstra{\ss}e 1, D-85748 Garching, Germany}

%% file: authors.tex
\author[0009-0001-5949-1524]{Hannah B. Koziol}
\affiliation{\UCSD}
\email{hkoziol@ucsd.edu}

\author[0000-0002-4378-8534]{Karin Sandstrom}
\affiliation{\UCSD}
\email{kmsandstrom@ucsd.edu}

\author[0000-0003-4974-3481]{Dalya Baron}
\affiliation{\Carnegie}
\affiliation{\KIPAC}
\email{dalyabaron@gmail.com}

\author[0000-0002-2545-1700]{Adam K. Leroy}
\affiliation{\OSU}
\affiliation{\Columbus}
\email{leroy.42@osu.edu}

\author[0000-0002-5235-5589]{Jérémy Chastenet}
\affiliation{\UGent}
\email{jchastenet@ucsd.edu}

\author[0000-0001-8241-7704]{Ryan Chown}
\affiliation{\Algoma}
\email{rchown53@gmail.com}

\author[0000-0002-5782-9093]{Daniel A. Dale}
\affiliation{\UWyoming}
\email{ddale@uwyo.edu}

\author[0000-0002-4755-118X]{Oleg V. Egorov}
\affiliation{\UHeidelberg}
\email{oleg.egorov@uni-heidelberg.de}

\author[0009-0005-0750-2956]{Lindsey Hands}
\affiliation{\UCSD}
\email{lhands@ucsd.edu}

\author[0000-0002-3472-0490]{Mansi Padave}
\affiliation{\UCSD}
\email{mpadave@ucsd.edu}

\author[0000-0003-2721-487X]{Debosmita Pathak}
\affiliation{\OSU}
\affiliation{\Columbus}
\email{pathak.89@buckeyemail.osu.edu}

\author[0000-0002-5204-2259]{Erik Rosolowsky}
\affiliation{\UAlberta}
\email{rosolowsky@ualberta.ca}

\author[0000-0002-9183-8102]{Jessica Sutter}
\affiliation{\Whitman}
\email{jessica.sutter93@gmail.com}

\author[0009-0005-8923-558X]{Tony D. Weinbeck}
\affiliation{\UWyoming}
\email{tonyweinbeck@gmail.com}

\author[0000-0002-0012-2142]{Thomas G. Williams}
\affiliation{\JBCA}
\email{thomas.g.williams@manchester.ac.uk}

\author[0000-0002-5480-5686]{Alberto Bolatto}
\affiliation{\umd}
\email{bolatto@umd.edu}

\author[0000-0003-0946-6176]{Médéric Boquien}
\affiliation{\Azur}
\email{mederic.boquien@oca.eu}

\author[0000-0001-5301-1326]{Yixian Cao}
\affiliation{\MPE}
\email{ycao@mpe.mpg.de}

\author[0000-0002-8549-4083]{Enrico Congiu}
\affiliation{\ESOcl}
\email{enry91con@gmail.com}

\author[0000-0001-6708-1317]{Simon C. O. Glover}
\affiliation{\ITA}
\email{glover@uni-heidelberg.de}

\author[0000-0003-4770-688X]{Hwihyun Kim}
\affiliation{\NOIRLab}
\email{hwihyun.kim@noirlab.edu}

\author[0000-0002-0560-3172]{Ralf S.\ Klessen}
\affiliation{\ITA}
\affiliation{\IWR}
\email{klessen@uni-heidelberg.de}

\author[0000-0003-3917-6460]{Kirsten~L.~Larson}
\STScIESA
\email{kilarson@stsci.edu}

\author[0000-0002-3289-8914]{Justus Neumann}
\affiliation{Max-Planck-Institut f\"{u}r Astronomie, K\"{o}nigstuhl 17, D-69117 Heidelberg, Germany}
\email{jneumann@mpia.de}

\author[0000-0002-0119-1115]{Elias K. Oakes}
\affiliation{\UConn}
\email{eliaskoakes@gmail.com}

\author[0000-0002-1370-6964]{Hsi-An Pan}
\affiliation{\Tamkang}
\email{hapan@gms.tku.edu.tw}

\author[0000-0002-0873-5744]{Ismael Pessa}
\affiliation{\AIP}
\email{ipessa@aip.de}

\author[0000-0002-6313-4597]{Sumit K. Sarbadhicary}
\JHU
\email{ssarbad1@jh.edu}

\author[0009-0000-2764-6723]{Benjamin W. Stadel}
\affiliation{\UAlberta}
\email{benstadel@shaw.ca}

\author[0000-0003-0378-4667]{Jiayi~Sun}
\affiliation{Department of Physics and Astronomy, University of Kentucky, 506 Library Drive, Lexington, KY 40506, USA}
\email{jiayi.sun@uky.edu}

\author[0000-0002-8528-7340]{David~A.~Thilker}
\JHU
\email{dthilker@jhu.edu}

\author[0000-0001-7130-2880]{Leonardo \'Ubeda}
\affiliation{\STScI}
\email{lubeda@stsci.edu}

%% file: table_phangs_sample.tex
\begin{deluxetable*}{lrrrrrccrr}
\tabletypesize{\small}
\tablecaption{PHANGS Cycle 1 JWST Sample \label{tab:table}}
\tablehead{
\colhead{Galaxy} &
\colhead{Distance} &
\colhead{P.A.} &
\colhead{i} &
\colhead{\newrefedit{r$_{25}$}} &
\colhead{Area} &
\colhead{$\log_{10}\Sigma_{M*}$} &
\colhead{$\log_{10}\Sigma_{\rm SFR}$} &
\colhead{\refedit{B$_{\rm{PAH}}$} Slope} &
\colhead{\refedit{B$_{\rm{PAH}}$} Slope} \\
\colhead{} &
\colhead{(Mpc)} &
\colhead{(deg)} &
\colhead{(deg)} &
\colhead{(\newrefedit{arcmin})} &
\colhead{(kpc$^{2}$)} &
\colhead{(M$_\odot$ kpc$^{-2}$)} &
\colhead{(M$_\odot$ yr$^{-1}$ kpc$^{-2}$)} &
\colhead{Equation \ref{s23}} &
\colhead{Equation \ref{eq_ssfr}} }
\startdata
IC 5332   & 9.01 & 74  & 27 & \newrefedit{3.03} & 198.61   & 7.4 & $-2.7$ & $1.77 \pm 0.0070$ & $1.71 \pm 0.15$ \\
NGC 0628  & 9.84 & 21  & 9  & \newrefedit{4.94} & 628.83   & 7.5 & $-2.6$ & $1.73 \pm 0.0012$ & $1.73 \pm 0.14$ \\
NGC 1087  & 15.85 & 359 & 43 & \newrefedit{1.49}  & 147.44  & 7.8 & $-2.1$ & $1.83 \pm 0.0011$ & $1.85 \pm 0.14$ \\
NGC 1300  & 18.99 & 278 & 32 & \newrefedit{2.97} & 846.43  & 7.7 & $-2.9$ & $1.70 \pm 0.0027$ & $1.68 \pm 0.15$ \\
NGC 1365  & 19.57 & 201 & 55 & \newrefedit{6.01} & 3678.9  & 7.4 & $-2.3$ & $1.71 \pm 0.0010$ & $1.70 \pm 0.15$ \\
NGC 1385  & 17.22 & 181 & 44 & \newrefedit{1.70}  & 228.35  & 7.6 & $-2.0$ & $1.91 \pm 0.0081$ & $1.88 \pm 0.14$ \\
NGC 1433  & 18.63 & 200 & 29 & \newrefedit{3.10} & 885.05  & 7.9 & $-2.9$ & $1.74 \pm 0.0024$ & $1.62 \pm 0.15$ \\
NGC 1512  & 18.83 & 262 & 43 & \newrefedit{4.22} & 1675.9  & 7.5 & $-3.1$ & $1.75 \pm 0.0026$ & $1.65 \pm 0.15$ \\
NGC 1566  & 17.69 & 215 & 30 & \newrefedit{3.61} & 1086.4  & 7.7 & $-2.4$ & $1.76 \pm 0.00071$ & $1.75 \pm 0.14$ \\
NGC 1672  & 19.40 & 134 & 43 & \newrefedit{3.08} & 946.56  & 7.8 & $-2.1$ & $1.85 \pm 0.00041$ & $1.81 \pm 0.14$ \\
NGC 2835  & 12.22 & 1   & 41 & \newrefedit{3.21} & 408.02  & 7.4 & $-2.5$ & $1.96 \pm 0.0023$ & $1.78 \pm 0.14$ \\
NGC 3351  & 9.96 & 193 & 45 & \newrefedit{3.61} & 344.40   & 7.8 & $-2.4$ & $1.60 \pm 0.0034$ & $1.70 \pm 0.15$ \\
NGC 3627  & 11.32 & 173 & 57 & \newrefedit{5.14} & 899.99  & 7.9 & $-2.4$ & $1.78 \pm 0.00047$ & $1.78 \pm 0.14$ \\
NGC 4254  & 13.10 & 68  & 34 & \newrefedit{2.52}  & 289.12  & 7.9 & $-2.0$ & $1.84 \pm 0.00066$ & $1.82 \pm 0.14$ \\
NGC 4303  & 16.99 & 312 & 24 & \newrefedit{3.44} & 909.77  & 7.6 & $-2.2$ & $1.80 \pm 0.00065$ & $1.83 \pm 0.14$ \\
NGC 4321  & 15.21 & 156 & 39 & \newrefedit{3.05} & 571.22  & 8.0 & $-2.2$ & $1.72 \pm 0.0010$ & $1.75 \pm 0.14$ \\
NGC 4535  & 15.77 & 180 & 45 & \newrefedit{4.07} & 1097.0  & 7.5 & $-2.7$ & $1.73 \pm 0.0026$ & $1.72 \pm 0.14$ \\
NGC 5068  & 5.20 & 342 & 36 & \newrefedit{3.74}  & 100.59   & 7.4 & $-2.6$ & $1.94 \pm 0.0011$ & $1.80 \pm 0.14$ \\
NGC 7496  & 18.72 & 194 & 36 & \newrefedit{1.67}  & 261.31  & 7.6 & $-2.1$ & $1.76 \pm 0.0028$ & $1.75 \pm 0.14$ \\
\enddata
\tablecomments{Distance measurements from the compilation by \citet{Anand2021}. Position angles (P.A.) and inclinations (i) from \citet{Lang2020} where available and \citet{Leroy2021} otherwise. Radii measured in r$_{25}$ from HyperLEDA database presented in \citet{Makarov2014}. Galaxy areas are calculated in kpc$^2$ using the r$_{25}$ and distance assuming a circular geometry. Surface densities of stellar mass, $\log_{10}\Sigma_{M*}$, from \citet{Leroy2021}. Surface densities of SFRs, $\log_{10}\Sigma_{\rm SFR}$, are measured from extinction-corrected H$\alpha$ maps from \citet{Belfiore2023}. \refedit{B$_{\rm{PAH}}$ s}lopes measured using the method described in Section \ref{sec:method} are listed under \refedit{B$_{\rm{PAH}}$} Slope Equation \ref{s23}. \refedit{B$_{\rm{PAH}}$ s}lopes calculated using Equation \ref{eq_ssfr} with global sSFR values are listed under \refedit{B$_{\rm{PAH}}$} Slope Equation \ref{eq_ssfr}.}
\end{deluxetable*}

%% file: table_correlations_spear.tex
\begin{deluxetable}{ccc}
\tabletypesize{\small}
\tablecaption{Correlation coefficients and $p$-values for various galaxy properties compared to B$_{\rm PAH}$ slope values calculated in each 1.5 kpc region, where y is the B$_{\rm PAH}$ slope value and x is the property. \label{tab:correlations}}
\tablehead{
\colhead{Property} & 
\colhead{Spearman $\rho$} & 
\colhead{$p$}
}
\startdata
log(sSFR) &  0.38 & $\ll 0.03$ \\
12$+$log(O/H) &  -0.28 & $\ll 0.03$ \\
log(R$_{\rm gal}$) & 0.047 & 0.90 \\
log($\Sigma_{\rm mol}/\Sigma_{\rm gas}$)   &  0.25 & 0.062 \\
log([OIII]/H$\beta$) &  -0.012 & 0.89 \\
log([NII]/H$\alpha$) &  -0.30  & $\ll 0.03$ \\
log([SII]/H$\alpha$) &  -0.048 & 0.89 \\
log(W4/W1)  &  0.18 & $\ll 0.03$ \\ 
\enddata
\end{deluxetable}

%% file: table_correlations_band_ratios_spear.tex
\begin{deluxetable}{cc}
\tablecaption{Correlation coefficients and $p$-values between PAH band ratios and physical properties in 1.5~kpc regions where \refedit{y} is the PAH band ratio and \refedit{x} is the property. \label{tab:band_ratios}}
\tabletypesize{\small}
\tablewidth{0.45\textwidth}
\tablehead{
\colhead{\textbf{3.3/11.3}} &
\colhead{\textbf{3.3/7.7}}
}
\startdata
\begin{tabular}{lc}
\hline
Property & Spearman ($\rho$, $p$) \\
\hline
log(sSFR) & 0.081 (0.86) \\
12 + log(O/H) &  -0.31 ($\ll$0.03) \\
log(R$_{\rm gal}$) & 0.20 ($\ll$0.03) \\
log($\Sigma_{\rm mol}/\Sigma_{\rm gas}$) &  0.11 (0.092) \\
log([OIII]/H$\beta$) &  -0.21 (0.10) \\
log([NII]/H$\alpha$) &  -0.36 ($\ll$0.03) \\
log([SII]/H$\alpha$) &  0.11 (0.041) \\
\hline
\end{tabular}
&
\begin{tabular}{c}
\hline
Spearman ($\rho$, $p$) \\
\hline
-0.20 ($\ll$0.03) \\
-0.20 ($\ll$0.03) \\
0.16 ($\ll$0.03) \\
-0.016 (0.92) \\
0.054 (0.82) \\
-0.11 (0.81) \\
0.40 ($\ll$0.03) \\
\hline
\end{tabular}
\enddata
\end{deluxetable}